\documentclass[aps,preprint]{revtex4}%
\usepackage{amsfonts}
\usepackage{amsmath}
\usepackage{amssymb}
\usepackage{graphicx}%
\providecommand{\U}[1]{\protect\rule{.1in}{.1in}}

\begin{document}
\preprint{ }
\title[Short title for running header]{Non-compact groups, Coherent States, Relativistic Wave equations and the
Harmonic Oscillator.}
\author{Diego Julio Cirilo-Lombardo}
\affiliation{Bogoliubov Laboratory of Theoretical Physics}
\affiliation{Joint Institute for Nuclear Research, 141980, Dubna, Russian Federation.}
\email{diego@thsun1.jinr.ru ; diego77jcl@yahoo.com}
\author{}

\begin{abstract}
Relativistic geometrical action for a quantum particle in the superspace is
analyzed from theoretical group point of view. To this end an alternative
technique of quantization outlined by the authors in a previous work and that
is based in the correct interpretation of the square root Hamiltonian, is
used. The obtained spectrum of physical states and the Fock construction
consist of Squeezed States which correspond to the representations with the
lowest weights $\lambda=\frac{1}{4}$ and $\lambda=\frac{3}{4}$ with four
possible (non-trivial) fractional representations for the group decomposition
of the spin structure. From the theory of semi-groups the analytical
representation of the radical operator in the superspace is constructed, the
conserved currents are computed and a new relativistic wave equation is
proposed and explicitly solved for the time dependent case. The relation with
the Relativistic Schr\"{o}dinger equation and the Time-dependent Harmonic
Oscillator is analyzed and discussed.


\end{abstract}
\volumeyear{year}
\volumenumber{number}
\issuenumber{number}
\eid{identifier}
\date[Date text]{date}
\received[Received text]{date}

\revised[Revised text]{date}

\accepted[Accepted text]{date}

\published[Published text]{date}

\startpage{0}
\endpage{ }
\maketitle
\tableofcontents

\section{Introduction and summary}

The quantum behaviour of a relativistic particle in the superspace, besides to
be a useful tool for certain studies and applications of Quantum Field Theory
(QFT), is of notable importance in many physical contexts. Time-dependent
Landau systems and the electron-monopole system are described naturally by the
Super-Heisenberg-Weyl and OSP(1/2) algebras [1-3]. If several more or less
well known physical systems are intrinsically supersymmetric in nature an
obvious following question was: Can any supersymmetric toy model give us a
good picture of not so well known physical systems? Part of the purpose of
this paper is to demonstrate the positive answer to this question showing that
a relativistic particle in the superspace can describe particles with
fractionary spin for the which not concrete action is known.

On the other hand, the Time-Dependent Harmonic Oscillator (TDHO) was
demonstrated to be a powerful tool to describe systems with more complicated
dynamics in closed form. From the famous reports of Ermakov [4] and Husimi [5]
we can see that if \ any physical problem with a complicated or envolved
dynamics can be represented faithfully or "mapped" to a TDHO system, this
complicated dynamics admits a Coherent State (CS) or Squeezed States (SS)
realization [6]. It is clearly important that the model proposed here admits a
Coherent State and Squeezed State realization. Most notably, squeezed states
have been used in the context of quantum optics [7] and in the context of
gravitational wave detection [8]. The correct choice for the realization of
the physical states, however, will depend on the symmetry group that defines
in some meaning the particular physical system under study. Another part of
this work will be devoted to discuss this point and what happens when
different algebras can characterize the same physical problem.

In a previous paper we considered the simple model of superparticle of Volkov
and Pashnev [9], that is type G4 in the description of Casalbuoni [10,11]$,$
in order to quantize it and to obtain the spectrum of physical states with the
Hamiltonian remaining in the natural square root form. To this end, we used
the Hamiltonian formulation described by Lanczos in [12] and the inhomogeneous
Lorentz group as a representation for the obtained physical states [13,14,15].
The quantization of this model was performed completely and the obtained
spectrum of physical states, with the Hamiltonian operator in its square root
form, was compared with the spectrum obtained with the Hamiltonian in the
standard form (i.e.: quadratic in momenta). We showed that the square root
Hamiltonian can operate in a natural manner with the states with correspond to
the representations with the lowest weights $\lambda_{1,2}=\frac{1}{4}$ and
$\lambda_{1,2}=\frac{3}{4}$ and that there are four possible (non-trivial)
fractional representations for the group decomposition of the spin structure
from the square root Hamiltonian, instead of (1/2,0) and (0,1/2), as the case
when the Hamiltonian is quadratic in momentum (e.g. Ref.[9]). This result was
a consequence of the geometrical Hamiltonian taken in its natural square root
form and the Sannikov-Dirac oscillator representation for the generators of
the Lorentz group SO(3,1). In this manner, we also showed that the
superparticle relativistic actions as of Ref.[9] are a good geometrical and
natural candidate to describe quartionic states [16,17,18] (semions). In this
paper we will complete the previous work giving now in more explicit form, how
the states can be faithfully represented and realized from the geometrical
point of view and from the dynamics of the group manifold.

In this work, strongly motivated for the several fundamental reasons described
above, we considered the same simple model of superparticle of Volkov and
Pashnev [9] to find the link with the TDHO problem, and for instance, with the
CS\ and SS representations of the physical states obtained in [19] with the
Hamiltonian remaining in its square root form. The plan of this paper is as
follows: in order to make this work self-contained, in Sections 2, 3 and 4 we
borrow from reference [19] the geometrical description, the Hamiltonian
treatment and quantization of the superparticle model. Section 5 is devoted to
describe the process of quantization and the obtaining of the mass spectrum of
the superparticle model under consideration emphasizing the relation between
the group representation of the physical states and their CS or SS
realizations. In Section 6 from the theory of semigroups we construct a
general analytic representation of the square root Hamiltonian. In Sections 7
and 8 the relation of the model with the relativistic Schr\"{o}dinger equation
is discussed and a new relativistic wave equation is proposed. Finally, some
conclusions and remarks are given in Section 9.

\section{The superparticle model}

\noindent In the superspace the coordinates are given not only by the
space-time $x_{\mu}$ coordinates, but also for anticommuting spinors
$\theta^{\alpha}$ and $\overline{\theta}^{\overset{.}{\alpha}}$ . The
resulting metric [9,19,20] must be invariant to the action of the Poincare
group, and also invariant to the supersymmetry transformations
\[
x_{\mu}^{\prime}=x_{\mu}+i\left(  \theta^{\alpha}\left(  \sigma\right)
_{\alpha\overset{.}{\beta}}\overline{\xi}^{\overset{.}{\beta}}-\xi^{\alpha
}\left(  \sigma\right)  _{\alpha\overset{.}{\beta}}\overline{\theta}%
^{\overset{.}{\beta}}\right)  \ ,\ \ \theta^{\prime\alpha}=\theta^{\alpha}%
+\xi^{\alpha}\ ,\ \ \overline{\theta^{\prime}}^{\overset{.}{\alpha}}%
=\overline{\theta}^{\overset{.}{\alpha}}+\overline{\xi}^{\overset{.}{\alpha}}%
\]

The simplest super-interval that obeys the requirements of invariance given
above, is the following (our choice for the metric tensor is $g_{\mu\nu
}=(+---)$)
\begin{equation}
ds^{2}=\omega^{\mu}\omega_{\mu}+a\omega^{\alpha}\omega_{\alpha}-a^{\ast}%
\omega^{\overset{.}{\alpha}}\omega_{\overset{.}{\alpha}} \tag{1}%
\end{equation}
where (to simplify notation from here we avoid the contracted indexes between
the spin-tensors $\left(  \sigma\right)  _{\alpha\overset{.}{\beta}}$ and the
anticommuting spinors $\theta^{\alpha}$ and $\overline{\theta}^{\overset
{.}{\alpha}}$, as usual)
\[
\omega_{\mu}=dx_{\mu}-i\left(  d\theta\ \sigma_{\mu}\overline{\theta}%
-\theta\ \sigma_{\mu}d\overline{\theta}\right)  ,\ \ \ \ \ \ \ \ \omega
^{\alpha}=d\theta^{\alpha}\ ,\ \ \ \ \ \ \ \ \omega^{\overset{.}{\alpha}%
}=\overline{d\theta}^{\overset{.}{\alpha}}%
\]
are the Cartan forms of the group of supersymmetry [20].

The spinorial indexes are related as follows
\[
\theta^{\alpha}=\varepsilon^{\alpha\beta}\theta_{\beta}%
\ \ \ \ \ ,\ \ \ \ \ \theta_{\alpha}=\theta^{\beta}\varepsilon_{\beta\alpha
}\ \ \ ,\ \ \varepsilon_{\alpha\beta}=-\varepsilon_{\beta\alpha}%
\ \ \ \ ,\ \ \ \varepsilon^{\alpha\beta}=-\varepsilon^{\beta\alpha
}\ \ \ ,\ \ \ \varepsilon_{12}=\varepsilon^{12}=1
\]
and of analog manner for the spinors with punctuated indexes. The complex
constants $a$ and $a^{\ast}$ in the line element (1) are arbitrary. This
arbitrarity for the choice of $a$ and $a^{\ast}$are constrained by the
invariance and reality of the interval (1).

As we have extended our manifold to include fermionic coordinates, it is
natural to extend also the concept of trajectory of point particle to the
superspace. To do this we take the coordinates $x\left(  \tau\right)  $,
$\theta^{\alpha}\left(  \tau\right)  $ and $\overline{\theta}^{\overset
{.}{\alpha}}\left(  \tau\right)  $ depending on the evolution parameter
$\tau.$ Geometrically, the function action that will describe the world-line
of the superparticle, is
\begin{equation}
S=-m\int_{\tau1}^{\tau2}d\tau\sqrt{\overset{\circ}{\omega_{\mu}}\overset
{\circ}{\omega^{\mu}}+a\overset{.}{\theta}^{\alpha}\overset{.}{\theta}%
_{\alpha}-a^{\ast}\overset{.}{\overline{\theta}}^{\overset{.}{\alpha}}%
\overset{.}{\overline{\theta}}_{\overset{.}{\alpha}}}=\int_{\tau1}^{\tau
2}d\tau L\left(  x,\theta,\overline{\theta}\right)  \tag{2}%
\end{equation}
where $\overset{\circ}{\omega_{\mu}}=\overset{.}{x}_{\mu}-i\left(  \overset
{.}{\theta}\ \sigma_{\mu}\overline{\theta}-\theta\ \sigma_{\mu}\overset
{.}{\overline{\theta}}\right)  $ and the upper point means derivative with
respect to the parameter $\tau$, as usual.

The momenta, canonically conjugated to the coordinates of the superparticle,
are
\[
\mathcal{P}_{\mu}=\partial L/\partial x^{\mu}=\left(  m^{2}/L\right)
\overset{\circ}{\omega_{\mu}}%
\]%
\[
\mathcal{P}_{\alpha}=\partial L/\overset{.}{\partial\theta^{\alpha}%
}=i\mathcal{P}_{\mu}\left(  \sigma^{\mu}\right)  _{\alpha\overset{.}{\beta}%
}\overline{\theta}^{\overset{.}{\beta}}+\left(  m^{2}a/L\right)  \overset
{.}{\theta_{\alpha}}%
\]%
\begin{equation}
\mathcal{P}_{\overset{.}{\alpha}}=\partial L/\overset{.}{\partial
\overline{\theta}^{\overset{.}{\alpha}}}=i\mathcal{P}_{\mu}\theta^{\alpha
}\left(  \sigma^{\mu}\right)  _{\alpha\overset{.}{\alpha}}-\left(
m^{2}a/L\right)  \overset{.}{\overline{\theta}_{\overset{.}{\alpha}}} \tag{3}%
\end{equation}
It is difficult to study this system in the Hamiltonian formalism framework
because of the constraints and the nullification of the Hamiltonian. As the
action (2)\ is invariant under reparametrizations of the evolution parameter
\[
\tau\rightarrow\widetilde{\tau}=f\left(  \tau\right)
\]
one way to overcome this difficulty is to make the dynamic variable $x_{0}$
the time. For this, it is sufficient to introduce the concept of integration
and derivation in supermanifolds as we introduce in Section 6\footnotetext[1]%
{We take the Berezin convention for the Grassmannian derivatives: $\ \delta
F(\theta)=\frac{\partial F}{\partial\theta}\delta\theta$} to write the action
in the form
\[
S=-m\int_{\tau1}^{\tau2}\overset{.}{x}_{0}d\tau\sqrt{\left[  1-iW_{,0}%
^{0}\right]  ^{2}-\left[  x_{,0}^{i}-W_{,0}^{i}\right]  ^{2}+\overset{.}%
{x}_{0}^{-2}\left(  a\overset{.}{\theta_{\alpha}}\overset{.}{\theta^{\alpha}%
}-a^{\ast}\overset{.}{\overline{\theta}_{\overset{.}{\alpha}}}\overset
{.}{\overline{\theta}^{\overset{.}{\alpha}}}\right)  }%
\]
where the $W_{,0}^{\mu}$ was defined by
\[
\overset{\circ}{\omega}^{0}=\overset{.}{x}^{0}\left[  1-iW_{,0}^{0}\right]
\]%
\[
\overset{\circ}{\omega}^{i}=\overset{.}{x}^{0}\left[  x_{,0}^{i}-iW_{,0}%
^{i}\right]
\]
whence $x_{0}\left(  \tau\right)  $ turns out to be the evolution parameter
\[
S=-m\int_{x_{0}\left(  \tau_{1}\right)  }^{x_{0}\left(  \tau_{2}\right)
}dx_{0}\sqrt{\left[  1-iW_{,0}^{0}\right]  ^{2}-\left[  x_{,0}^{i}-W_{,0}%
^{i}\right]  ^{2}+a\overset{.}{\theta}^{\alpha}\overset{.}{\theta}_{\alpha
}-a^{\ast}\overset{.}{\overline{\theta}}^{\overset{.}{\alpha}}\overset
{.}{\overline{\theta}}_{\overset{.}{\alpha}}}\equiv\int dx_{0}L
\]
Physically, this parameter (we call it the dynamical parameter) is the time
measured by an observer's clock in the rest frame.

Therefore, the invariance of a theory with respect to the invariance of the
coordinate evolution parameter means that one of the dynamic variables of the
theory ($x_{0}\left(  \tau\right)  $ in this case) becomes the observed time
with the corresponding non-zero Hamiltonian
\begin{equation}
H=\mathcal{P}_{\mu}\overset{.}{x}^{\mu}+\Pi^{\alpha}\overset{.}{\theta
}_{\alpha}+\Pi^{\overset{.}{\alpha}}\overset{.}{\theta}_{\overset{.}{\alpha}%
}-L\nonumber
\end{equation}%
\begin{equation}
\ \ \ \ \ \ \ \ \ \ \ \ \ \ \ \ \ =\sqrt{m^{2}-\left(  \mathcal{P}%
_{i}\mathcal{P}^{i}+\frac{1}{a}\Pi^{\alpha}\Pi_{\alpha}-\frac{1}{a^{\ast}}%
\Pi^{\overset{.}{\alpha}}\Pi_{\overset{.}{\alpha}}\right)  } \tag{4}%
\end{equation}
where
\[
\Pi_{\alpha}=\mathcal{P}_{\alpha}+i\ \mathcal{P}_{\mu}\left(  \sigma^{\mu
}\right)  _{\alpha\overset{.}{\beta}}\overline{\theta}^{\overset{.}{\beta}}%
\]%
\[
\Pi_{\overset{.}{\alpha}}=\mathcal{P}_{\overset{.}{\alpha}}-i\mathcal{P}_{\mu
}\theta^{\alpha}\left(  \sigma^{\mu}\right)  _{\alpha\overset{.}{\alpha}}%
\]
That gives the well known mass shell condition and losing, from the quantum
point of view, the operatibility of the Hamiltonian.

In the paper[9], where this type of superparticle action was explicitly
presented, the problem of nullification of Hamiltonian was avoided in the
standard form. This means that the analog to a mass shell condition (4) in
superspace was introduced by means of a multiplier (einbein) to obtain a new
Hamiltonian
\begin{equation}
H=\frac{\varkappa}{2}\left\{  m^{2}-\mathcal{P}_{0}\mathcal{P}^{0}-\left(
\mathcal{P}_{i}\mathcal{P}^{i}+\frac{1}{a}\Pi^{\alpha}\Pi_{\alpha}-\frac
{1}{a^{\ast}}\Pi^{\overset{.}{\alpha}}\Pi_{\overset{.}{\alpha}}\right)
\right\}  \tag{5}%
\end{equation}
(here $\varkappa,$ as in Ref. [9], is a constant with the dimensions of the
square of a length).With this Hamiltonian it is clear that in order to perform
the quantization of the superparticle the problems disappear: $\mathcal{P}%
_{0}\,$is restored into the new Hamiltonian, and the square root is
eliminated. The full spectrum from this Hamiltonian was obtained in [9] where
the quantum Hamiltonian referred to the center of mass was
\begin{equation}
H_{cm}=\frac{\varkappa}{2}\left\{  m^{2}-M^{2}+\frac{2^{3/2}M}{\left\vert
a\right\vert }\left[  1-\left(  \sigma_{0}\right)  _{\alpha\overset{.}{\beta}%
}\overline{s}^{\overset{.}{\beta}}s^{\alpha}\right]  \right\}  \tag{6a}%
\end{equation}
with the mass distribution of the physical states being the following : two
scalar supermultiplets $M_{1s}=\frac{2^{1/2}}{\left\vert a\right\vert }$
$+\sqrt{\frac{2}{\left\vert a\right\vert }+m^{2}}$ and $M_{2s}=\sqrt{\frac
{2}{\left\vert a\right\vert }+m^{2}}-\frac{2^{1/2}}{\left\vert a\right\vert }%
$; and one vector supermultiplet $M_{v}=m$. The Fock's construction in the
center of mass for the eq.(6a) ( Hamiltonian quadratic in momenta) consists of
the following vectors:
\begin{equation}%
\begin{array}
[c]{lllll}%
S_{1}=\left\vert 0\right\rangle e^{iMt} &  & \Xi_{1\alpha}=\overline
{d}_{\overset{\cdot}{\alpha}}\left\vert 0\right\rangle e^{iMt} &  &
P_{1}=\overline{d}^{\overset{\cdot}{\beta}}\overline{d}_{\overset{\cdot}%
{\beta}}\left\vert 0\right\rangle e^{iMt}\\
&  &  &  & \\
\Xi_{2\alpha}=\overline{s}_{\overset{\cdot}{\alpha}}\left\vert 0\right\rangle
e^{iMt} &  & V_{\alpha\beta}=\overline{s}_{\overset{\cdot}{\alpha}}%
\overline{d}_{\overset{\cdot}{\beta}}\left\vert 0\right\rangle e^{iMt} &  &
\Xi_{3\alpha}=\overline{s}_{\overset{\cdot}{\alpha}}\overline{d}%
^{\overset{\cdot}{\beta}}\overline{d}_{\overset{\cdot}{\beta}}\left\vert
0\right\rangle e^{iMt}\\
&  &  &  & \\
P_{2}=\overline{s}^{\overset{\cdot}{\alpha}}\overline{s}_{\overset{\cdot
}{\alpha}}\left\vert 0\right\rangle e^{iMt} &  & \Xi_{4\alpha}=\overline
{d}_{\overset{\cdot}{\alpha}}\overline{s}^{\overset{\cdot}{\beta}}\overline
{s}_{\overset{\cdot}{\beta}}\left\vert 0\right\rangle e^{iMt} &  & \\
&  &  &  & \\
S_{2}=\overline{d}^{\overset{\cdot}{\beta}}\overline{d}_{\overset{\cdot}%
{\beta}}\overline{s}^{\overset{\cdot}{\alpha}}\overline{s}_{\overset{\cdot
}{\alpha}}\left\vert 0\right\rangle e^{iMt} &  &  &  &
\end{array}
\tag{6b}%
\end{equation}
where operators $s_{\alpha}$ and $d_{\alpha}$ acting on the vacuum give zero:
$s_{\alpha}\left\vert 0\right\rangle =$ $d_{\alpha}\left\vert 0\right\rangle
=0$.

We will show in this report that it is possible, in order to quantize the
superparticle action, to remain the Hamiltonian in the square root form. As it
is very obvious, in the form of square root the Hamiltonian operator is not
linearly proportional with the operator $n_{s}=\overline{s}^{\overset{.}%
{\beta}}s^{\alpha}$. The Fock construction for the Hamiltonian into the square
root form agrees formally with the description given above for reference[9],
but the operability of this Hamiltonian is over basic states with lowest
helicities $\lambda=1/4,3/4$. This means that the superparticle Hamiltonian
preserving the square root form operates over physical states of particles
with fractional quantum statistics and fractional spin (quartions).

\section{Hamiltonian treatment in Lanczo's formulation}

In order to solve our problem from the dynamical and quantum mechanical point
of view, we will use the formulation given in[12,21]. This Hamiltonian
formulation for dynamical systems was proposed by C. Lanczos and allows us to
preserve the square root form in the new Hamiltonian. We start from expression
(4)
\[
H=\sqrt{m^{2}-\left(  \mathcal{P}_{i}\mathcal{P}^{i}+\frac{1}{a}\Pi^{\alpha
}\Pi_{\alpha}-\frac{1}{a^{\ast}}\Pi^{\overset{.}{\alpha}}\Pi_{\overset
{.}{\alpha}}\right)  }%
\]
if
\[
\frac{dt}{d\tau}\equiv\frac{dx^{0}}{d\tau}=g\left(  \mathcal{P}_{0}%
,\ \mathcal{P}_{i},\ \Pi_{\alpha},\ \Pi_{\overset{.}{\alpha}},\ x_{0}%
,\ x_{i},\ \theta_{\alpha},\ \overline{\theta}_{\overset{.}{\alpha}}\right)
\]
with the arbitrary function $g$ given by
\begin{equation}
g=\frac{\left(  \sqrt{m^{2}-\left(  \mathcal{P}_{i}\mathcal{P}^{i}+\frac{1}%
{a}\Pi^{\alpha}\Pi_{\alpha}-\frac{1}{a^{\ast}}\Pi^{\overset{.}{\alpha}}%
\Pi_{\overset{.}{\alpha}}\right)  }-\mathcal{P}_{0}\right)  \sqrt
{m^{2}-\mathcal{P}_{0}\mathcal{P}^{0}-\left(  \mathcal{P}_{i}\mathcal{P}%
^{i}+\frac{1}{a}\Pi^{\alpha}\Pi_{\alpha}-\frac{1}{a^{\ast}}\Pi^{\overset
{.}{\alpha}}\Pi_{\overset{.}{\alpha}}\right)  }}{m^{2}-\left(  \mathcal{P}%
_{i}\mathcal{P}^{i}+\frac{1}{a}\Pi^{\alpha}\Pi_{\alpha}-\frac{1}{a^{\ast}}%
\Pi^{\overset{.}{\alpha}}\Pi_{\overset{.}{\alpha}}\right)  -\mathcal{P}%
_{0}^{2}} \tag{7}%
\end{equation}
the new Hamiltonian $\mathcal{H}$ takes the required \textquotedblright square
root\textquotedblright\ form
\begin{equation}
\mathcal{H}\equiv g\left(  H+\mathcal{P}_{0}\right)  =\sqrt{m^{2}%
-\mathcal{P}_{0}\mathcal{P}^{0}-\left(  \mathcal{P}_{i}\mathcal{P}^{i}%
+\frac{1}{a}\Pi^{\alpha}\Pi_{\alpha}-\frac{1}{a^{\ast}}\Pi^{\overset{.}%
{\alpha}}\Pi_{\overset{.}{\alpha}}\right)  } \tag{8}%
\end{equation}
where we shall set $\mathcal{H}=0$ (now depending on $2n+2$ canonical
variables), and the variable $\mathcal{P}_{0}$ is clearly identificated by the
dynamical expression
\begin{equation}
\frac{d\mathcal{P}_{0}}{d\tau}=-g\frac{\partial\mathcal{H}}{\partial x^{0}%
}\ \text{ \ \ or \ \ \ }\frac{d\mathcal{P}_{0}}{d\tau}=-\frac{\partial
\mathcal{H}}{\partial t} \tag{9}%
\end{equation}
This means that $\ \mathcal{P}_{0}=-H+const.$ Concerning the equations of
motion , the integration and derivatives are consistent with the geometrical
treatment of supermanifolds that we will describe with some detail in Section
6 .

In order to make an analysis of the dynamics of our problem, we can compute
the Poisson brackets between all the canonical variables and their conjugate
momentum [9,10,11]
\begin{equation}
\overset{\cdot}{\mathcal{P}}_{\mu}=\left\{  \mathcal{P}_{\mu},\mathcal{H}%
\right\}  _{pb}=0 \tag{10}%
\end{equation}%
\begin{equation}
\overset{.}{\theta}^{\alpha}=\left\{  \theta^{\alpha},\mathcal{H}\right\}
_{pb}=\frac{1}{a}\frac{\Pi^{\alpha}}{\mathcal{H}} \tag{11}%
\end{equation}%
\begin{equation}
\overset{.}{\overline{\theta}}^{\overset{\cdot}{\alpha}}=\left\{
\overline{\theta}^{\overset{\cdot}{\alpha}},\mathcal{H}\right\}  _{pb}%
=-\frac{1}{a^{\ast}}\frac{\Pi^{\overset{\cdot}{\alpha}}}{\mathcal{H}} \tag{12}%
\end{equation}%
\begin{equation}
\overset{\cdot}{x}_{\mu}=\left\{  x_{\mu},\mathcal{H}\right\}  _{pb}=\frac
{1}{\mathcal{H}}\left\{  \mathcal{P}_{\mu}+\frac{i}{a}\ \Pi^{\alpha}%
(\sigma_{\mu})_{\alpha\overset{.}{\beta}}\overline{\theta}^{\overset{.}{\beta
}}+\frac{i}{a^{\ast}}\theta^{\alpha}(\sigma_{\mu})_{\alpha\overset{.}{\beta}%
}\Pi^{\overset{\cdot}{\beta}}\right\}  \tag{13}%
\end{equation}%
\begin{equation}
\overset{\cdot}{\Pi}_{\alpha}=\left\{  \Pi_{\alpha},\mathcal{H}\right\}
_{pb}=\frac{2i}{a^{\ast}\mathcal{H}}\mathcal{P}_{\alpha\overset{.}{\beta}}%
\Pi^{\overset{\cdot}{\beta}} \tag{14}%
\end{equation}%
\begin{equation}
\overset{\cdot}{\Pi}_{\overset{\cdot}{\alpha}}=\left\{  \Pi_{\overset{\cdot
}{\alpha}\ },\mathcal{H}\right\}  _{pb}=\frac{-2i}{a\mathcal{H}}\Pi^{\beta
}\mathcal{P}_{\beta\overset{.}{\alpha}} \tag{15}%
\end{equation}
where $\mathcal{P}_{\alpha\overset{.}{\beta}}\equiv\mathcal{P}_{\mu}\left(
\sigma^{\mu}\right)  _{\alpha\overset{.}{\beta}}$ . From the above expressions
the set of classical equations to solve is easily seen
\begin{equation}
\overset{\cdot\cdot}{\Pi}_{\alpha}=-\left(  \frac{4\mathcal{P}^{2}}{\left\vert
a\right\vert ^{2}\mathcal{H}^{2}}\right)  \overset{\cdot}{\Pi}_{\alpha}
\tag{16}%
\end{equation}%
\begin{equation}
\overset{\cdot\cdot}{\Pi}_{\overset{\cdot}{\alpha}}=-\left(  \frac
{4\mathcal{P}^{2}}{\left\vert a\right\vert ^{2}\mathcal{H}^{2}}\right)
\overset{\cdot}{\Pi}_{\overset{.}{\alpha}} \tag{17}%
\end{equation}
Assigning $\frac{4\mathcal{P}^{2}}{\left\vert a\right\vert ^{2}\mathcal{H}%
^{2}}\equiv\omega^{2}$, and having account for$\ \Pi_{\alpha}^{+}%
=-\Pi_{\overset{\cdot}{\alpha}}$ , the solution to the equations (16) and (17)
takes the form
\[
\Pi_{\alpha}=\xi_{\alpha}\ e^{i\omega\tau}+\eta_{\alpha}\ e^{-i\omega\tau}%
\]%
\begin{equation}
\Pi_{\overset{\cdot}{\alpha}}=-\overline{\eta}_{\overset{\cdot}{\alpha}%
}\ e^{i\omega\tau}-\overline{\xi}_{\overset{\cdot}{\alpha}}\ e^{-i\omega\tau}
\tag{18}%
\end{equation}
By means of the substitution of above solutions into (14) and (15), we find
the relation between $\xi_{\alpha}$ and $\eta_{\alpha}$%
\[
\eta_{\alpha}=\left(  \frac{2}{a^{\ast}\mathcal{H}\omega}\right)
\ \mathcal{P}_{\alpha\overset{.}{\beta}}\overline{\xi}^{\overset{.}{\beta}}%
\]
From eqs. (18) and above we obtain
\begin{equation}
\Pi_{\alpha}=\xi_{\alpha}\ e^{i\omega\tau}+\left(  \frac{2}{a^{\ast
}\mathcal{H}\omega}\right)  \ \mathcal{P}_{\alpha\overset{.}{\beta}}%
\overline{\xi}^{\overset{.}{\beta}}\ e^{-i\omega\tau} \tag{19}%
\end{equation}%
\begin{equation}
\Pi_{\overset{\cdot}{\alpha}}=-\left(  \frac{2}{a\mathcal{H}\omega}\right)
\ \xi^{\beta}\mathcal{P}_{\beta\overset{.}{\alpha}}\ e^{i\omega\tau}%
-\overline{\xi}_{\overset{\cdot}{\alpha}}\ e^{-i\omega\tau} \tag{20}%
\end{equation}
where we used the fact that the constant two-component spinors $\xi_{\alpha}$
verify $\overline{\xi}_{\overset{\cdot}{\alpha}}=\xi_{\alpha}^{+}$ .
Integrating expressions (11) and (12), we obtain explicitly the following
\begin{equation}
\theta_{\alpha}=\zeta_{\alpha}-\frac{i}{a\mathcal{H}\omega}\left[  \xi
_{\alpha}\ e^{i\omega\tau}-\frac{2}{a^{\ast}\mathcal{H}\omega}\ \mathcal{P}%
_{\alpha\overset{.}{\beta}}\overline{\xi}^{\overset{.}{\beta}}\ e^{-i\omega
\tau}\right]  \tag{21}%
\end{equation}%
\begin{equation}
\ \overline{\theta}_{\overset{.}{\alpha}}=\overline{\zeta}_{\overset{\cdot
}{\alpha}}+\frac{i}{a^{\ast}\mathcal{H}\omega}\left[  -\frac{2}{a\mathcal{H}%
\omega}\ \xi^{\beta}\mathcal{P}_{\beta\overset{.}{\alpha}}\ e^{i\omega\tau
}+\overline{\xi}_{\overset{\cdot}{\alpha}}\ e^{-i\omega\tau}\right]  \tag{22}%
\end{equation}
where $\zeta_{\alpha}$ and $\overline{\zeta}_{\overset{\cdot}{\alpha}}%
=\zeta_{\alpha}^{+}$ are two-component constant spinors.

Analogically, from expression (13), we obtain $x_{\mu}$ in explicit form
\begin{align}
x_{\mu}  &  =q_{\mu}-\frac{1}{\mathcal{H}}\left[  \mathcal{P}_{\mu}%
-\frac{\omega\mathcal{H}}{\mathcal{P}^{2}}\left(  \xi\sigma_{\mu}\overline
{\xi}\right)  \right]  \tau+\frac{1}{\mathcal{H}\omega}\left[  \frac{1}%
{a}\ e^{i\omega\tau}\left(  \xi\sigma_{\mu}\overline{\zeta}\right)  +\frac
{1}{a^{\ast}}\ \ e^{-i\omega\tau}\left(  \zeta\sigma_{\mu}\overline{\xi
}\right)  \right]  +\tag{23}\\
&  +\frac{\mathcal{P}_{\mu}}{2\mathcal{P}^{2}}\left[  \zeta^{\alpha}%
\xi_{\alpha}e^{i\omega\tau}-\overline{\zeta}^{\overset{\cdot}{\alpha}%
}\overline{\xi}_{\overset{\cdot}{\alpha}}\ e^{-i\omega\tau}\right] \nonumber
\end{align}

\section{Quantization}

Because of the correspondence between classical and quantum dynamics, the
Poisson brackets between coordinates and canonical impulses are transformed
into quantum commutators and anti-commutators
\[
\left[  x_{\mu},\mathcal{P}_{\mu}\right]  =i\left\{  x_{\mu},\mathcal{P}_{\mu
}\right\}  _{pb}=-ig_{\mu\nu}%
\]%
\[
\left\{  \theta^{\alpha},\mathcal{P}_{\beta}\right\}  =i\left\{
\theta^{\alpha},\mathcal{P}_{\beta}\right\}  _{pb}=-i\delta_{\beta
}^{\ \ \alpha}%
\]%
\begin{equation}
\left\{  \theta^{\overset{\cdot}{\alpha}},\mathcal{P}_{\overset{\cdot}{\beta}%
}\right\}  =i\left\{  \theta^{\overset{\cdot}{\alpha}},\mathcal{P}%
_{\overset{\cdot}{\beta}}\right\}  _{pb}=-i\delta_{\overset{\cdot}{\beta}%
}^{\ \ \overset{\cdot}{\alpha}} \tag{24}%
\end{equation}
and the new Hamiltonian (8) operates quantically as follows
\begin{equation}
\sqrt{m^{2}-\mathcal{P}_{0}\mathcal{P}^{0}-\left(  \mathcal{P}_{i}%
\mathcal{P}^{i}+\frac{1}{a}\Pi^{\alpha}\Pi_{\alpha}-\frac{1}{a^{\ast}}%
\Pi^{\overset{.}{\alpha}}\Pi_{\overset{.}{\alpha}}\right)  }\left\vert
\Psi\right\rangle =0 \tag{25}%
\end{equation}
where $\left\vert \Psi\right\rangle $ are the physical states. From the
(anti)commutation relations (24) it is possible to obtain easily the
commutators between the variables $\xi_{\alpha},$ $\overline{\xi}%
_{\overset{\cdot}{\alpha}},$ $\zeta_{\alpha},\overline{\text{ }\zeta
}_{\overset{\cdot}{\alpha}},$ $q_{\mu},$ $\mathcal{P}_{\mu}$%
\begin{equation}%
\begin{array}
[c]{lllll}%
\left\{  \xi_{\alpha},\overline{\xi}_{\overset{\cdot}{\alpha}}\right\}
=-\mathcal{P}_{\alpha\overset{.}{\alpha}} &  & \left\{  \zeta_{\alpha
},\overline{\zeta}_{\overset{\cdot}{\alpha}}\right\}  =-\left(  \frac
{1}{2\mathcal{P}^{2}}\right)  \mathcal{P}_{\alpha\overset{.}{\alpha}} &  &
\left[  q_{\mu},\mathcal{P}_{\mu}\right]  =-ig_{\mu\nu}%
\end{array}
\tag{26}%
\end{equation}
To obtain the physical spectrum we use the relations given by (26) into (25)
and the Hamiltonian $\mathcal{H}$ takes the following form
\begin{equation}
\mathcal{H}=\sqrt{m^{2}-\mathcal{P}_{0}\mathcal{P}^{0}-\mathcal{P}%
_{i}\mathcal{P}^{i}-\frac{2^{3/2}\sqrt{(\mathcal{P}_{\mu})^{2}}}{\left\vert
a\right\vert }-\frac{2^{3/2}}{\left\vert a\right\vert \sqrt{(\mathcal{P}_{\mu
})^{2}}}\xi^{\alpha}\mathcal{P}_{\alpha\overset{.}{\beta}}\overline{\xi
}^{\overset{.}{\beta}}\ } \tag{27}%
\end{equation}
Passing to the center of mass of the system, and defining new operators
$s_{\alpha}=(1/\sqrt{M})\xi_{\alpha}$, $\overline{s}_{\overset{.}{\alpha}%
}=(1/\sqrt{M})\overline{\xi}_{\overset{.}{\alpha}}$, $d_{\alpha}=\sqrt
{2M}\zeta_{\alpha}$, $\overline{d}_{\overset{.}{\alpha}}=\sqrt{2M}%
\overline{\zeta}_{\overset{.}{\alpha}}$ (where $M=\mathcal{P}_{0}$), the
$\mathcal{H}_{cm}$ is
\begin{equation}
\mathcal{H}_{cm}=\sqrt{m^{2}-M^{2}+\frac{2^{3/2}M}{\left\vert a\right\vert
}\left[  1-\left(  \sigma_{0}\right)  _{\alpha\overset{.}{\beta}}\overline
{s}^{\overset{.}{\beta}}s^{\alpha}\right]  \ } \tag{28}%
\end{equation}
being the anti-commutation relations of the operators $s_{\alpha},\overline
{s}_{\overset{.}{\alpha}},$ $d_{\alpha}$, $\overline{d}_{\overset{.}{\alpha}%
}:$
\begin{equation}%
\begin{array}
[c]{lll}%
\left\{  s_{\alpha},\overline{s}_{\overset{\cdot}{\alpha}}\right\}  =-\left(
\sigma_{0}\right)  _{\alpha\overset{.}{\alpha}} &  & \left\{  d_{\alpha
},\overline{d}_{\overset{\cdot}{\alpha}}\right\}  =-\left(  \sigma_{0}\right)
_{\alpha\overset{.}{\alpha}}%
\end{array}
\tag{29}%
\end{equation}
Now the question is: how does the square-root $\mathcal{H}$ Hamiltonian given
by expression (28) operate on a given physical state? The problem of locality
and interpretation of the operator like (25) is very well known. Several
attempts to avoid these problems were given in the literature [22,23]. The
main characteristic of all these attempts is to eliminate the square root of
the Hamiltonian: e.g.imposing constraints. In this manner, the set of
operators into the square root operates freely on the physical states, paying
the price to lose the concept of locality and quantum interpretation of the
spectrum of a well possed field theory.

Our plan is "to take " the square root to a bispinor in order to introduce the
physical state into the square root Hamiltonian. In the next section we will
perform the square root of a bispinor and obtain the mass spectrum given by
the Hamiltonian $\mathcal{H}$.

\section{Mass spectrum and square root of a bispinor}

The square root from a spinor was extracted in 1965 by the soviet scientist S.
S. Sannikov from Kharkov (Ukraine) [14] and the analysis of the structure of
the Hilbert space containing such "square root " states was worked out by E.
C. G. Sudarshan, N. Mukunda and C. C. Chiang in 1981 [24]. Taking the square
root from a spinor was performed also by P.A.M. Dirac [15] in 1971.

We know that the group $SL(2,\mathbb{C})$ is locally isomorph to SO(3,1), and
SL(2,R) is locally isomorph to SO(2,1). For instance, the generators of the
group SO(3,1) for our case can be constructed from the usual operators $a$,
$a^{+}$ (or $q$ and $p$) in the following manner: we start from an irreducible
unitary infinite dimensional representation of the Heisenberg-Weyl group,
which is realized in the Fock space of states of one-dimensional quantum
oscillator [13,17,18]. Creation operators and annihilation operators of these
states obey the conventional commutation relations $%
\begin{array}
[c]{lll}%
\left[  a,a^{+}\right]  =1 &  & \left[  a,a\right]  =\left[  a^{+}%
,a^{+}\right]  =0
\end{array}
.$ To describe this representation to the Lorentz group one may also use the
coordinate-momentum realization ($q,p=-i\frac{\partial}{\partial q}$) of the
Heisenberg algebra, which relates to the $a,a^{+}$ realization by the
formulas
\begin{equation}%
\begin{array}
[c]{lll}%
a=\frac{q+ip}{\sqrt{2}} &  & a^{+}=\frac{q-ip}{\sqrt{2}}%
\end{array}
\tag{30}%
\end{equation}
as usual. Let us introduce the spinors
\begin{equation}%
\begin{array}
[c]{lll}%
L_{\alpha}=\left(
\begin{array}
[c]{l}%
a_{1}\\
a_{1}^{+}%
\end{array}
\right)  &  & L_{\overset{\cdot}{\alpha}}=\left(
\begin{array}
[c]{l}%
a_{2}\\
a_{2}^{+}%
\end{array}
\right)
\end{array}
\tag{31}%
\end{equation}
The commutation relations take the form
\begin{equation}%
\begin{array}
[c]{lll}%
\left[  L_{\alpha},L_{\beta}\right]  =i\varepsilon_{\alpha\beta}\ ; & \left[
L_{\overset{\cdot}{\alpha}},L_{\overset{\cdot}{\beta}}\right]  =i\varepsilon
_{\overset{\cdot}{\alpha}\overset{\cdot}{\beta}}\ ; & \left[  L_{\overset
{\cdot}{\alpha}},L_{\beta}\right]  =0
\end{array}
\tag{32}%
\end{equation}
The generators of $SL(2,\mathbb{C})$ are easily constructed [18] from
$L_{\alpha}$ and $L_{\overset{\cdot}{\alpha}}$
\[
S_{\alpha\beta}\equiv iS_{1i}(\sigma^{i})_{\alpha\beta}=\frac{1}{4}\left\{
L_{\alpha},L_{\beta}\right\}
\]%
\begin{equation}
S_{\overset{\cdot}{\alpha}\overset{\cdot}{\beta}}\equiv iS_{2i}(\sigma
^{i})_{\overset{\cdot}{\alpha}\overset{\cdot}{\beta}}=\frac{1}{4}\left\{
L_{\overset{\cdot}{\alpha}},L_{\overset{\cdot}{\beta}}\right\}  \tag{33}%
\end{equation}
where the index $i=1,2,3$ and the six vectors $S_{ai}$ $\left(  a,b=1,2;a\neq
b\right)  $, characteristics of the representation of $SL(2,\mathbb{C}%
)\approx$ $SO(3,1)$, satisfy the commutation relations
\begin{equation}%
\begin{array}
[c]{lll}%
\left[  S_{ai},S_{aj}\right]  =-i\varepsilon_{ijk}S_{a}^{k}\ , & \left[
S_{bi},S_{bj}\right]  =-i\varepsilon_{ijk}S_{b}^{k}\ , & \left[  S_{ai}%
,S_{bj}\right]  =0
\end{array}
\tag{34}%
\end{equation}
Notice that the above construction \ obeys the described decomposition of
$SL(2,\mathbb{C})\approx$ $SO(3,1)$

Then the quantities
\begin{equation}%
\begin{array}
[c]{lll}%
\Phi_{\alpha}\equiv\left\langle \Psi\right\vert L_{\alpha}\left\vert
\Psi\right\rangle  & , & \ \ \ \ \ \ \ \ \ \ \overline{\Phi}_{\overset{\cdot
}{\alpha}}\equiv\left\langle \overline{\Psi}\right\vert L_{\overset{\cdot
}{\alpha}}\left\vert \overline{\Psi}\right\rangle ,
\end{array}
\tag{35}%
\end{equation}
are the two-components of a bispinor
\[
\Phi\equiv\left\langle \widehat{\Psi}\right\vert L\left\vert \widehat{\Psi
}\right\rangle =\left(
\begin{array}
[c]{l}%
\Phi_{\alpha}\\
\overline{\Phi}_{\overset{\cdot}{\alpha}}%
\end{array}
\right)
\]
where we define $\left\vert \widehat{\Psi}\right\rangle $ $\equiv\left(
\begin{array}
[c]{l}%
\left\vert \Psi\right\rangle \\
\left\vert \overline{\Psi}\right\rangle
\end{array}
\right)  $. Notice that $\left\vert \Psi\right\rangle $ and $\left\vert
\overline{\Psi}\right\rangle $ are the square root of each component of this
bispinor and can have the same form (given the isomorphism between the
generators $L_{\alpha}$ and $L_{\overset{\cdot}{\alpha}}$), that is very easy
to verify. In terms of $q$ the basic vectors of the representation can be
written as [13,14,17]%

\begin{equation}
\left\langle q\right.  \left\vert n\right\rangle =\varphi_{n}\left(  q\right)
=\pi^{-1/4}\left(  2^{n}n!\right)  ^{-1/2}H_{n}\left(  q\right)  e^{-q^{2}/2}
\tag{36}%
\end{equation}%
\begin{equation}
\int dq\varphi_{m}^{\ast}\left(  q\right)  \varphi_{n}\left(  q\right)
=\delta_{mn} \tag{37}%
\end{equation}
(where $H_{n}\left(  q\right)  $ are the Hermite polynomials) and form a
unitary representation of $SO\left(  3,1\right)  $, and
\begin{equation}
\left\vert n\right\rangle =\left(  n!\right)  ^{-1/2}\left(  a^{+}\right)
^{n}\left\vert 0\right\rangle \tag{38}%
\end{equation}
the normalized basic states where the vacuum vector is annihilated by $a$ .
The Casimir operator, that is $S_{ai}S_{a}^{i}$, has the eigenvalue
$\lambda(\lambda-1)=-\frac{3}{16}$ (for each subgroup ISO(2,1) given by
eqs.(33))and indeed corresponds to the representations with the lowest weights
$\lambda=\frac{1}{4}$ and $\lambda=\frac{3}{4}$ . The wave functions which
transform as linear irreducible representation of ISO$\left(  2,1\right)  $ ,
subgroup of ISO (3,1) generated by operators (33) are
\begin{equation}
\Psi_{1/4}\left(  x,\theta,q\right)  =\overset{+\infty}{\underset{k=0}{\sum}%
}f_{2k}\left(  x,\theta\right)  \varphi_{2k}\left(  q\right)  \tag{39}%
\end{equation}%
\begin{equation}
\Psi_{\ 3/4}\left(  x,\theta,q\right)  =\overset{+\infty}{\underset{k=0}{\sum
}}f_{2k+1}\left(  x,\theta\right)  \varphi_{2k+1}\left(  q\right)  \tag{40}%
\end{equation}
(analogically for the $\overline{\Psi}_{1/4}$ and $\overline{\Psi}_{3/4}$
states with contrary helicity). We can easily see that the Hamiltonian
$\mathcal{H}$ (28) operates over the states $\left\vert \widehat{\Psi
}\right\rangle $, which become into $\mathcal{H}$ as its square $\Phi_{\alpha
}$ and $\Phi_{\overset{.}{\alpha}}$. It is natural to associate, up to a
proportional factor, the spinors $d_{\alpha}$ and $\overline{d}_{\overset
{\cdot}{\alpha}}$ with
\begin{equation}%
\begin{array}
[c]{lll}%
d_{\alpha}\rightarrow\left(  \Phi_{1/4}\right)  _{\alpha}\equiv\left\langle
\Psi_{\ 1/4}\right\vert \mathbb{L}_{\alpha}\left\vert \Psi_{\ 1/4}%
\right\rangle  &  & \overline{d}_{\overset{\cdot}{\alpha}}\rightarrow\left(
\overline{\Phi}_{1/4}\right)  _{\overset{\cdot}{\alpha}}\equiv\left\langle
\overline{\Psi}_{1/4}\right\vert \mathbb{L}_{\overset{\cdot}{\alpha}%
}\left\vert \overline{\Psi}_{1/4}\right\rangle
\end{array}
\tag{41}%
\end{equation}
and, analogically, the spinors $s_{\alpha}$ and $\overline{s}_{\overset{\cdot
}{\alpha}}$ with
\begin{equation}%
\begin{array}
[c]{lll}%
s_{\alpha}\rightarrow\left(  \Phi_{3/4}\right)  _{\alpha}\equiv\left\langle
\Psi_{\ 3/4}\right\vert \mathbb{L}_{\alpha}\left\vert \Psi_{\ 3/4}%
\right\rangle  &  & \overline{s}_{\overset{\cdot}{\alpha}}\rightarrow\left(
\overline{\Phi}_{3/4}\right)  _{\overset{\cdot}{\alpha}}\equiv\left\langle
\overline{\Psi}_{\ 3/4}\right\vert \mathbb{L}_{\overset{\cdot}{\alpha}%
}\left\vert \overline{\Psi}_{\ 3/4}\right\rangle
\end{array}
\tag{42}%
\end{equation}
where the new spinors $\mathbb{L}_{\alpha}\left(  \mathbb{L}_{\overset{\cdot
}{\alpha}}\right)  $ are defined as%
\begin{equation}
\mathbb{L}_{\alpha}=\left(
\begin{array}
[c]{l}%
a_{1}a_{1}\\
a_{1}^{+}a_{1}^{+}%
\end{array}
\right)  \tag{42a}%
\end{equation}%
\[
\mathbb{L}_{\overset{.}{\alpha}}=\left(
\begin{array}
[c]{l}%
a_{2}a_{2}\\
a_{2}^{+}a_{2}^{+}%
\end{array}
\right)
\]
The reason to this choice is the following: as it was shown in Ref. [25] the
Hilbert space for each subgroup ISO(2,1)$\approx SU(1,1)$ [26] can be
decomposed as direct sum of two independent subspaces characterized for the
states of helicity $\lambda=\frac{1}{4}$ and $\lambda=\frac{3}{4}$
respectively. Each subspace is composed by the even$\left(  \lambda=\frac
{1}{4}\right)  $ and odd states $\left(  \lambda=\frac{3}{4}\right)  $ given
by eqs.(39-40). These "cat" states admit (after a convenient choice for the
functions $f_{2k}\left(  x,\theta\right)  $ and $f_{2k+1}\left(
x,\theta\right)  $) a coherent state realization being eigenvectors not of the
ladder operator $a$ of the Heisenberg-Weyl algebra , but those of the
quadratic ladder operator $aa$ of the $SU(1,1)$ algebra defined in general by
\begin{equation}%
\begin{array}
[c]{ccc}%
K_{+}=\frac{1}{2}a^{+}a^{+}\ , & K_{-}=\frac{1}{2}aa\ , & K_{0}=\frac{1}%
{4}\left(  a^{+}a+aa^{+}\right)  \ ,
\end{array}
\tag{42b}%
\end{equation}
That means that when we are in the full Hilbert space the algebra is
Heisenberg-Weyl and the states $\left\vert \Psi\right\rangle =\frac{1}%
{\sqrt{2}}\left(  \left\vert \Psi_{\ 1/4}\right\rangle +\left\vert
\Psi_{\ 3/4}\right\rangle \right)  $are eigenvectors of the operator $a,$ and
when we pass to the decomposed space (by means of a suitable unitary
transformation) the algebra becomes the $SU(1,1)$ algebra with the quadratic
ladder operators given by expression (42b).

The relations (41) and (42) give a natural link between the spinors
$\xi_{\alpha}\left(  \overline{\xi}_{\overset{\cdot}{\alpha}}\right)  $and
$\zeta_{\alpha}\left(  \overline{\zeta}_{\overset{\cdot}{\alpha}}\right)  $,
solutions of the dynamical problem, with the only physical states that can
operate freely with the Hamiltonian $\mathcal{H}$ : the \textquotedblright
square root\textquotedblright\ states $\left\vert \Psi\right\rangle $,
$\left\vert \overline{\Psi}\right\rangle $ from the bispinor $\Phi.$ Notice
that there are four (non-trivial) representations for the group decomposition
of the bispinor $\Phi,$as follows
\[
\Phi_{1}=\left(
\begin{array}
[c]{l}%
\Phi_{\ 1/4}\\
\overline{\Phi}_{\ 3/4}%
\end{array}
\right)  \rightarrow\left(  1/4,0\right)  \oplus\left(  0,3/4\right)
\]%
\[
\Phi_{2}=\left(
\begin{array}
[c]{l}%
\Phi_{\ 3/4}\\
\overline{\Phi}_{\ 1/4}%
\end{array}
\right)  \rightarrow\left(  3/4,0\right)  \oplus\left(  0,1/4\right)
\]%
\[
\Phi_{3}=\left(
\begin{array}
[c]{l}%
\Phi_{\ 1/4}\\
\overline{\Phi}_{\ 1/4}%
\end{array}
\right)  \rightarrow\left(  1/4,0\right)  \oplus\left(  0,1/4\right)
\]%
\[
\Phi_{4}=\left(
\begin{array}
[c]{l}%
\Phi_{\ 3/4}\\
\overline{\Phi}_{\ 3/4}%
\end{array}
\right)  \rightarrow\left(  3/4,0\right)  \oplus\left(  0,3/4\right)
\]
This result is a consequence of the geometrical Hamiltonian taken in its
natural square root form and the Sannikov-Dirac oscillator representation for
the generators of the Lorentz group SO(3,1).

Commutation relations (29) obey the Clifford's algebra for spinorial
creation-annihilation operators. In this manner, the square root of the
operators $s_{\alpha}$ and $d_{\alpha}$ in the representation given by the
associations (41) and (42) acting on the vacuum give zero, symbolically :
\[
\sqrt{s_{\alpha}}\rightarrow\ \left(  \Phi_{3/4}\right)  _{\alpha}\left\vert
0\right\rangle =\mathbb{L}_{\alpha}^{rs}\Psi_{r\ 3/4}\Psi_{s\ 3/4}\left\vert
0\right\rangle =0
\]%
\[
\sqrt{d_{\alpha}}\rightarrow\ \left(  \Phi_{1/4}\right)  _{\alpha}\left\vert
0\right\rangle =\mathbb{L}_{\alpha}^{rs}\Psi_{\ r\ 1/4}\Psi_{s\ 3/4}\left\vert
0\right\rangle =0
\]
where here we introduce $r,s,t...$ latin indexes to design the fractional spin
states. The Fock's construction in the center of mass of the system consists
now, in contrast to the construction (6b), of the following vectors:
\[%
\begin{array}
[c]{lll}%
\widehat{S}_{1}=\left\vert 0\right\rangle e^{\frac{iMt}{2}}, & \ \ \ \ \Xi
_{1r}=\overline{\Psi}_{r\ 1/4}\left\vert 0\right\rangle e^{\frac{iMt}{2}}, &
\ \ \ \ \widehat{P}_{1}=\overline{\Psi}_{1/4}^{r\ }\overline{\Psi}%
_{r\ 1/4}\left\vert 0\right\rangle e^{\frac{iMt}{2}},
\end{array}
\]%
\begin{equation}%
\begin{array}
[c]{lll}%
\Xi_{2r}=\overline{\Psi}_{r\ 3/4}\left\vert 0\right\rangle e^{\frac{iMt}{2}%
}, & \ \ \ \ V_{rs}=\overline{\Psi}_{r\ 3/4}\overline{\Psi}_{s\ 1/4}\left\vert
0\right\rangle e^{\frac{iMt}{2}}, & \ \ \ \ \Xi_{3r}=\overline{\Psi}%
_{r\ 3/4}\overline{\Psi}_{1/4}^{s\ }\overline{\Psi}_{s\ 1/4}\left\vert
0\right\rangle e^{\frac{iMt}{2}},
\end{array}
\tag{43}%
\end{equation}%
\[%
\begin{array}
[c]{ll}%
\widehat{P}_{2}=\overline{\Psi}_{3/4}^{r\ }\overline{\Psi}_{r\ 3/4}\left\vert
0\right\rangle e^{\frac{iMt}{2}}, & \ \ \ \ \Xi_{4r}=\overline{\Psi}%
_{r\ 1/4}\overline{\Psi}_{3/4}^{s\ }\overline{\Psi}_{s\ 3/4}\left\vert
0\right\rangle e^{\frac{iMt}{2}},
\end{array}
\]%
\[
\widehat{S}_{2}=\overline{\Psi}_{1/4}^{r\ }\overline{\Psi}_{r\ 1/4}%
\overline{\Psi}_{3/4}^{s\ }\overline{\Psi}_{s\ 3/4}\left\vert 0\right\rangle
e^{\frac{iMt}{2}}%
\]

Notice that the vectors given above are the only states that can operate into
the square root operator given by expression (28), and not that constructed
directly with the operators $s_{\alpha}$ and $d_{\alpha}$. Schematically we
have, e.g. for $\Xi_{4r}$, the following operability :
\[
\sqrt{m^{2}-M^{2}+\frac{2^{3/2}M}{\left\vert a\right\vert }\left[  1-\left(
\sigma_{0}\right)  _{\alpha\overset{.}{\beta}}\overline{s}^{\overset{.}{\beta
}}s^{\alpha}\right]  \ }\overline{\Psi}_{r\ 1/4}\overline{\Psi}_{3/4}%
^{s\ }\overline{\Psi}_{s\ 3/4}\left\vert 0\right\rangle e^{\frac{iMt}{2}%
}\equiv
\]%
\[
\equiv\sqrt{\left[  m^{2}-M^{2}+\frac{2^{3/2}M}{\left\vert a\right\vert
}\left[  1-\left(  \sigma_{0}\right)  _{\alpha\overset{.}{\beta}}\overline
{s}^{\overset{.}{\beta}}s^{\alpha}\right]  \right]  \overline{d}%
_{\overset{\cdot}{\gamma}}\overline{s}^{\overset{\cdot}{\beta}}\overline
{s}_{\overset{\cdot}{\beta}}\ e^{iMt}\ }\left\vert 0\right\rangle
\]
From expression (38) and taking into account that the number operator is
$\overline{s}^{\overset{.}{\beta}}s^{\alpha}\equiv n_{s}$, because
$\overline{s}^{\overset{.}{\beta}}$ and $s^{\alpha}$ work as
creation-annihilation operators, we can easily obtain the mass for the
different \textquotedblright square root\textquotedblright\ or fractional
supermultiplets :

$i)$ $n_{s}=0\rightarrow M_{I}=\sqrt{-\frac{2^{1/2}}{\left\vert a\right\vert
}+\sqrt{\frac{2}{\left\vert a\right\vert ^{2}}+m^{2}}}$ ; Fractional
supermultiplet I:$\left(  \widehat{S}_{1},\Xi_{1r},\widehat{P}_{1}\right)  $

$ii)$ $n_{s}=1\rightarrow M_{II}=\sqrt{m}$ ; Fractional supermultiplet II:
$\left(  \Xi_{2r},V_{rs},\Xi_{3r}\right)  $.

$iii)$ $n_{s}=2\rightarrow M_{III}=\sqrt{\sqrt{\frac{2}{\left\vert
a\right\vert ^{2}}+m^{2}}+\frac{2^{1/2}}{\left\vert a\right\vert }}$ ;
Fractional supermultiplet III: $\left(  \widehat{P}_{2},\Xi_{4r},\widehat
{S}_{2}\right)  $.

We emphasize now that the computations and algebraic manipulations given above
were with $\overline{d}_{\overset{\cdot}{\alpha}}\rightarrow\left(  \Phi
_{1/4}\right)  _{\overset{\cdot}{\alpha}}$ and $\overline{s}_{\overset{\cdot
}{\alpha}}\rightarrow\left(  \Phi_{3/4}\right)  _{\overset{\cdot}{\alpha}}$
(the square of the true states) into the square root Hamiltonian. That means
that the physical states are constrained by the the explicit form of the
Hamiltonian operator. Notice from expressions (35), (41) and (42) that the
physical states for the Hamiltonian in the square root form are one half the
number of physical states for the Hamiltonian quadratic in momenta.

Another important point is that the link between the new Hamiltonian
$\mathcal{H}$ given by expression (8) and the relativistic Schr\"{o}dinger
equation (e.g. ref.$^{27}$) can be given through the relation between the
conserved currents of the fermionic "square" states and the para-states. This
important issue will be analyzed in Section VII.

It is interesting to note that the arbitrary c-parameters $a$ and $a^{\ast}$
generate a deformation of the usual line element for a superparticle in proper
time, and this deformation is responsible, in any meaning, for the multiplets
given above. This is not a casuality: one can easily see how the quantum
Hamiltonian (28) is modified in the center of mass of the system by the
c-parameters $a$ and $a^{\ast\text{ }}$. The implications of this type of
superparticle actions with deformations of the quantization will be analyzed
in a future paper [28].

\section{Square root Hamiltonian and the Theory of Semigroups}

From the Hamiltonian eq.(8)
\[
\mathcal{H}=\sqrt{m^{2}-\mathcal{P}_{0}\mathcal{P}^{0}-\left(  \mathcal{P}%
_{i}\mathcal{P}^{i}+\frac{1}{a}\Pi^{\alpha}\Pi_{\alpha}-\frac{1}{a^{\ast}}%
\Pi^{\overset{.}{\alpha}}\Pi_{\overset{.}{\alpha}}\right)  }%
\]
where
\[
\Pi_{\alpha}=\mathcal{P}_{\alpha}+i\ \mathcal{P}_{\mu}\left(  \sigma^{\mu
}\right)  _{\alpha\overset{.}{\beta}}\overline{\theta}^{\overset{.}{\beta}}%
\]%
\[
\Pi_{\overset{.}{\alpha}}=\mathcal{P}_{\overset{.}{\alpha}}-i\mathcal{P}_{\mu
}\theta^{\alpha}\left(  \sigma^{\mu}\right)  _{\alpha\overset{.}{\alpha}}%
\]
($\mathcal{P}_{\alpha}$ and$\ \mathcal{P}_{\mu}$ was defined from the
Lagrangian, as usual), we start with the equation
\begin{equation}
\mathcal{S}\left[  \Psi\right]  =\mathcal{H}_{s}\Psi=\sqrt{m^{2}%
-\mathcal{P}_{0}\mathcal{P}^{0}-\left(  \mathcal{P}_{i}\mathcal{P}^{i}%
+\frac{1}{a}\Pi^{\alpha}\Pi_{\alpha}-\frac{1}{a^{\ast}}\Pi^{\overset{.}%
{\alpha}}\Pi_{\overset{.}{\alpha}}\right)  }\Psi\tag{44}%
\end{equation}
In order to construct a general analytic representation for above equation,
let us set $\omega^{2}=m^{2}$ and $-\mathcal{P}_{0}\mathcal{P}^{0}-\left(
\mathcal{P}_{i}\mathcal{P}^{i}+\frac{1}{a}\Pi^{\alpha}\Pi_{\alpha}-\frac
{1}{a^{\ast}}\Pi^{\overset{.}{\alpha}}\Pi_{\overset{.}{\alpha}}\right)
=-\mathbb{G}$.

Because we are treating it now with a supermanifold instead of the simple flat
space, it is necessary to give a consistent definition of integration on it,
in order to perform successfully the construction of the general analytical
representation of eq.(44). To do this we make firstly some remarks on
superspace and supermanifolds.

The supermanifolds we are dealing with are modelled over flat superspace
$B_{L}^{\left(  D_{0},D_{1}\right)  }$, the cartesian product of $D_{0}$
copies of $B_{L,0}$ and $D_{1}$ copies of $B_{L,1}$, where $B_{L,0}$ and
$B_{L,1}$ are the even, respectively odd, subspaces of a real Grassmann
algebra $B_{L}$ (with $L$ anticommuting generators). Functions from
$B_{L}^{\left(  D_{0},D_{1}\right)  }$ to $B_{L}$ will be taken to be
$G^{\infty}$, i.e. infinitely differentiable with respect to all arguments
which in turn implies that the function admits a finite Taylor series
expansion in the odd arguments, with infinitely differentiable functions of
the even arguments as coefficients [29]. A $\left(  D_{0}+D_{1}\right)
$-dimensional supermanifold $M^{\left(  D_{0},D_{1}\right)  }$ is constructed
from $B_{L}^{\left(  D_{0},D_{1}\right)  }$ in the usual way by means of an
atlas of charts $\underset{i\in I}{\cup}\left(  U_{i},\varphi_{i}\right)  $
with $U_{i}$ an open cover of $M^{\left(  D_{0},D_{1}\right)  }$ and a
homeomorphism $\varphi_{i}$ of $U_{i}$ onto an open subset of $B_{L}^{\left(
D_{0},D_{1}\right)  }$. If the overlap $U_{i}\cap U_{j}$ is non-empty we
require the transition function $\varphi_{j}\circ\varphi_{i}^{-1}$ to be
$G^{\infty}$.

A chart map $\varphi$ induces coordinates $\varphi^{M}\left(  m\right)
=z^{M}=\left\{  x^{\mu},\theta^{\alpha}\right\}  \left(  M=1,...,D_{0}%
+D_{1};\mu=1,...D_{0},\alpha=1,...,D_{1}\right)  $. If we change to other
coordinates we require this change to respect evenness/oddness in the
following sense: if $\left(  X\right)  $ denotes a grading of an Grassmann
element $X$, i.e. $\left(  X\right)  =0$ if $X$ is even and $\left(  X\right)
=1$ if $X$ is odd, then under a coordinate change $z^{M}\rightarrow
\overline{z}^{M}$ we require $\left(  z^{M}\right)  =\left(  \overline{z}%
^{M}\right)  $.

Equipped with the notion of differentiability in $B_{L}^{\left(  D_{0}%
,D_{1}\right)  }$ one can construct the tangent bundle $TM^{\left(
D_{0},D_{1}\right)  }$. At a point $m\in U\subset M^{\left(  D_{0}%
,D_{1}\right)  }$ the tangent space $\underset{\left(  m\right)  }%
{T}M^{\left(  D_{0},D_{1}\right)  }$ is spanned (on a coordinate basis) by
$\left\{  _{M}\partial=\frac{\overset{\rightarrow}{\partial}}{\partial z^{M}%
}\right\}  $(we use the de Witt conventions [30] for index manipulation in
order to avoid factors of $\left(  -1\right)  $). The dual space to
$\underset{\left(  m\right)  }{T}M^{\left(  D_{0},D_{1}\right)  }$ denoted
$\underset{\left(  m\right)  }{T^{\ast}}M^{\left(  D_{0},D_{1}\right)  }$ is
spanned by $\left\{  z^{M}\right\}  $ where $\left\langle _{N}\partial\right.
\left\vert z^{M}\right\rangle =\ _{N}\delta^{M}$ (The Kronecker delta). This
in turn gives rise to the cotangent bundle $T^{\ast}M$. In general, field
tensors of type $\left(  p,r\right)  $ are elements of $\otimes^{p}T^{\ast
}M^{\left(  D_{0},D_{1}\right)  }\otimes\otimes^{r}TM^{\left(  D_{0}%
,D_{1}\right)  }$, then the components of a tensor of type $\left(
p,r\right)  $ are displayed on a coordinate basis as%
\[
dz^{N_{1}}\otimes...\otimes dz^{N_{p}}\ \left(  _{N_{p...}N_{1}}%
T^{M_{1}....M_{r}}\right)  \ \otimes_{M_{r}}\partial\otimes...\otimes_{M_{1}%
}\partial
\]
We will need tensors with special symmetry properties that generalize the
differential forms of the ordinary differential geometry. This space of
$\left(  p,r\right)  $ tensors is denoted $\Lambda_{p}^{r}\left(  M^{\left(
D_{0},D_{1}\right)  }\right)  $ and spanned by%
\[
dz^{N_{1}}\wedge...\wedge dz^{N_{p}}\ \otimes\ _{M_{r}}\partial\vee
...\vee\ _{M_{1}}\partial
\]
where $\wedge$ is the graded antisymmetric wedge product%
\[
dz^{N}\wedge dz^{M}=dz^{(N}\wedge dz^{M]}=-\left(  -1\right)  ^{\left(
N\right)  \left(  M\right)  }dz^{M}\wedge dz^{N},
\]
where $\left(  N\right)  =\left(  dz^{N}\right)  $, and $\vee$ is the graded
symmetric product%
\[
\ _{N}\partial\vee\ _{M}\partial=\ _{(N}\partial\vee\ _{M]}\partial=+\left(
-1\right)  ^{\left(  N\right)  \left(  M\right)  }\ _{N}\partial\vee
\ _{M}\partial
\]
An element of $\Lambda_{p}^{r}\left(  M^{\left(  D_{0},D_{1}\right)  }\right)
$ is denominated a hyperform [31]. For $p=0$ a hyperform is also called a
derivative $r$-form; for $r=0$ a hyperform is a differential $p$-form.

We are now in the position to introduce the following definition that precises
the relationship between the supermanifold integration and the Berezin/Riemann integration:

\textit{Definition}. Let $M^{\left(  D_{0},D_{1}\right)  }$ be the total space
of a fibre bundle $\ E=(M^{\left(  D_{0},D_{1}\right)  },\pi,B_{0},F_{1}),$
and let $p\in M^{\left(  D_{0},D_{1}\right)  }$. Let $\left(  \pi^{-1}\left(
U\right)  ,\varphi\right)  $ be a chart on the $G^{\infty}$ supermanifold,
where $U\subset B_{0}$ and $p\in U$. Because of the local trivialization
property we take $\pi^{-1}\left(  U\right)  \cong U\times F_{1}$. Let $A$ be a
$\left(  D_{0}^{H},D_{1}^{V}\right)  $ hyperform with support compact in
$\varphi\left(  U\right)  \subset O$, where $O$ is open in $B_{L,0}^{D_{0}}$.
In natural coordinates $z^{M}=\varphi^{M}\left(  p\right)  $, and with respect
to the canonical basis%

\[
A=a\left(  z\right)  \Omega_{\left(  D_{0},D_{1}\right)  }\frac{1}{D_{0}%
!}\frac{1}{D_{1}!}dx^{\mu_{1}}\wedge...\wedge dx^{\mu_{D_{0}}}\ _{\mu
_{D_{0}...}\mu_{1}}A^{\alpha_{1}....\alpha_{D_{1}}}\ _{\alpha_{D_{1}}}%
\partial\vee...\vee\ _{\alpha_{1}}\partial
\]
Then%
\begin{equation}
\underset{\pi^{-1}(u)}{\int A}:=\overset{Ber}{\underset{O\times B_{L,1}%
^{D_{1}}}{\int}}a\left(  x,\theta\right)  dx^{1}...dx^{D_{0}}d\theta
^{1}...d\theta^{D_{1}} \tag{45}%
\end{equation}
Notice that it makes no sense to demand that $A$ has compact support in the "
$\theta$ direction", as e.g. for $B_{L}^{\left(  0,1\right)  }$ the only
$G^{\infty}$ function of $\theta$ which has compact support is ,, $f\left(
\theta\right)  =0$. By construction, the left-hand side of the definition (45)
transform with the Berezinian under the following change of coordinates%
\begin{equation}
\left\{  x^{\mu},\theta^{\alpha}\right\}  \rightarrow\left\{  \overline
{x}^{\mu}\left(  x\right)  ,\overline{\theta}^{\alpha}\left(  x,\theta\right)
\right\}  \tag{46}%
\end{equation}
that is nothing more than a bundle morphism from one set of natural
coordinates to another. This corresponds to the choice of a different section
and a change of basis in the fibres. From the practical point of view we can
easily see that, following the definition (45), the procedure consists of
replacing the one-forms $dx^{\mu}$ by the integration symbols
\textquotedblleft$dx^{\mu}$\textquotedblright, the derivative one-forms
$_{\alpha}\partial$ by the integration symbols \textquotedblleft%
$d\theta^{\alpha}$\textquotedblright, and deleting the $\wedge$\ and $\vee
$\ products. This procedure is justified by the fact that both sides transform
identically under the coordinate transformations (46).

We assume that $-\mathbb{G}+\omega^{2}$ satisfies the conditions required to
be a generator of a unitary group (self-adjoint); we can write (44) as :%

\begin{equation}
\mathcal{S}\left[  \Psi\right]  =\sqrt{-\mathbb{G}+\omega^{2}}\Psi\tag{47}%
\end{equation}
Using the analytic theory of fractional powers of closed linear operators
(e.g. Refs. [27,32]), it can be shown that (for generators of unitary groups)
we can write (47) as
\begin{equation}
\mathcal{S}\left[  \Psi\right]  =\frac{1}{\pi}\int_{0}^{\infty}\left[
-\mathbb{G}+\left(  \lambda+\omega^{2}\right)  \right]  ^{-1}\Psi
\frac{d\lambda}{\sqrt{\lambda}} \tag{48}%
\end{equation}
where $\left[  -\mathbb{G}+\left(  \lambda+\omega^{2}\right)  \right]  ^{-1}$
is the resolvent associated with the operator $\left(  -\mathbb{G}+\omega
^{2}\right)  .$ The resolvent can be computed directly if we can find the
fundamental solution to the equation
\begin{equation}
\frac{\partial Q}{\partial t}(\mathbf{z}_{1},\mathbf{z}_{2};t)+\left(
\mathbb{G}-\omega^{2}\right)  Q(\mathbf{z}_{1},\mathbf{z}_{2};t)=\delta
(\mathbf{z}_{1}-\mathbf{z}_{2}) \tag{49}%
\end{equation}
It is shown in [27] that the equation
\begin{equation}
i\hbar\frac{\partial\overline{Q}}{\partial t}(\mathbf{z}_{1},\mathbf{z}%
_{2};t)+\left(  \frac{1}{2M}\mathbb{G}-V\right)  \overline{Q}(\mathbf{z}%
_{1},\mathbf{z}_{2};t)=\delta(\mathbf{z}_{1}-\mathbf{z}_{2}) \tag{50}%
\end{equation}
has the general (infinitesimal) solution
\begin{equation}
\overline{Q}(\mathbf{z}_{1},\mathbf{z}_{2};t)=\left(  \frac{M}{2\pi i\hbar
t}\right)  ^{3/2}e^{\left\{  \frac{it}{\hbar}\left[  \frac{M}{2}\left(
\frac{\mathbf{z}_{1}-\mathbf{z}_{2}}{t}\right)  ^{2}-V\left(  z_{2}\right)
\right]  +\frac{ie}{\hbar c}(\mathbf{z}_{1}-\mathbf{z}_{2})\mathbb{A}\left[
\frac{1}{2}(\mathbf{z}_{1}-\mathbf{z}_{2})\right]  \right\}  } \tag{51}%
\end{equation}
provided that $\mathbb{A}$ and $M$ are time independent. We used the midpoint
evaluation in the last part for equation (51) $\left(  \mathbb{A}\left[
\frac{1}{2}(\mathbf{z}_{1}-\mathbf{z}_{2})\right]  \right)  $ and from the
supersymmetric Hamiltonian (44) that $\mathbb{A\equiv}\left(
0,-i\ \mathcal{P}_{\mu}\left(  \sigma^{\mu}\right)  _{\alpha\overset{.}{\beta
}}\overline{\theta}^{\overset{.}{\beta}},i\ \mathcal{P}_{\mu}\left(
\sigma^{\mu}\right)  _{\alpha\overset{.}{\beta}}\overline{\theta}^{\overset
{.}{\beta}}\right)  $ in superspace components.

If we set $\frac{\omega^{2}}{i\hbar}=V$ and $M=\frac{i\hbar}{2}$ then
\[
Q(\mathbf{z}_{1},t;\mathbf{z}_{2},0)=\int_{\mathbf{z}\left(  0\right)
=\mathbf{\ z}_{2}}^{\mathbf{z}\left(  t\right)  =\mathbf{z}_{1}}%
\mathcal{DW}_{\mathbf{z},t}[\mathbf{z}\left(  s\right)  ]\ e^{\left\{
\int_{0}^{t}V[\mathbf{z}\left(  s\right)  ]ds+\frac{ie}{\hbar c}%
\int_{\mathbf{y}}^{\mathbf{x}}\mathbb{A}[\mathbf{z}\left(  s\right)
]d\mathbf{z}\left(  s\right)  \right\}  }%
\]
solves (50), where
\[
\int_{\mathbf{z}\left(  0\right)  =\mathbf{\ z}_{2}}^{\mathbf{z}\left(
t\right)  =\mathbf{z}_{1}}\mathcal{DW}_{\mathbf{z},t}[\mathbf{z}\left(
s\right)  ]=\int_{\mathbf{z}\left(  0\right)  =\mathbf{\ z}_{2}}%
^{\mathbf{z}\left(  t\right)  =\mathbf{z}_{1}}\mathcal{D}[\mathbf{z}\left(
s\right)  ]\ e^{\left\{  -\frac{1}{4}\int_{0}^{t}\left\vert \frac
{d\mathbf{z}\left(  s\right)  }{ds}\right\vert ^{2}ds\right\}  }%
\]%
\[
=\underset{N\rightarrow\infty}{\lim}\left[  \frac{1}{4\pi\varepsilon\left(
N\right)  }\right]  ^{n\frac{N}{2}}%
{\displaystyle\int_{\mathcal{M}}}
\overset{N}{\underset{j=1}{\prod}}dz_{j}\ e^{\left\{  -\underset{j=1}%
{\overset{N}{\sum}}\left[  \frac{1}{4\varepsilon\left(  N\right)  }\left(
z_{j}-z_{j-1}\right)  ^{2}\right]  \right\}  }%
\]
and $\varepsilon\left(  N\right)  =t/N$.

Now we construct the solution of equation (47) for the constant $\mathbb{A}$
case. The solution for other examples will be discussed in a future paper
applying explicitly these results to the Hamiltonian formulation in
supermanifolds. First, rewrite equation (51) as%
\[
Q(\mathbf{z}_{1},\mathbf{z}_{2};t)=\left(  \frac{1}{4\pi t}\right)  ^{3/2}%
\exp\left\{  \frac{\left\Vert \left(  \mathbf{z}_{1}-\mathbf{z}_{2}\right)
\right\Vert ^{2}}{4s}-\mu^{2}t+\frac{ie}{\hbar c}\left(  \mathbf{z}%
_{1}-\mathbf{z}_{2}\right)  .\mathbb{A}\right\}
\]
Finally, using the theory of fractional powers, we note if $\mathbf{T}\left[
t,0\right]  $ is the semigroup associated with $-\mathbb{G}+\omega^{2}$, then
the semigroup associated with $\left[  -\mathbb{G}+\omega^{2}\right]  ^{1/2}$
is given by$^{27,32}$
\[
\mathbf{T}_{1/2}\left[  t,0\right]  \varphi\left(  \mathbf{x}\right)
=\int_{0}^{\infty}\left\{  \int_{\mathcal{M}}\left(  \frac{1}{4\pi t}\right)
^{3/2}\exp\left[  \frac{\left(  \mathbf{z}_{1}-\mathbf{z}_{2}\right)  ^{2}%
}{4t}-\mu^{2}t+\frac{ie}{\hbar c}\left(  \mathbf{z}_{1}-\mathbf{z}_{2}\right)
.\mathbb{A}\right]  \varphi\left(  \mathbf{z}_{2}\right)  d\mathbf{z}%
_{2}\right\}  \times
\]%
\[
\times\left(  \frac{ct}{\sqrt{4\pi}}\right)  \frac{1}{s^{3/2}}\exp\left(
-\frac{\left(  ct\right)  ^{2}}{4s}\right)  ds
\]
Laplace transforming we get%
\begin{equation}
\mathbf{T}_{1/2}\left[  t,0\right]  \varphi\left(  \mathbf{x}\right)
=\frac{ct}{4\pi^{2}}\int_{\mathcal{M}}\left(  \frac{1}{4\pi t}\right)
^{3/2}\exp\left[  \frac{ie}{2\hbar c}\left(  \mathbf{z}_{1}-\mathbf{z}%
_{2}\right)  .\mathbb{A}\right]  \frac{2\mu^{2}K_{2}\left[  \mu\left(
\left\Vert \left(  \mathbf{z}_{1}-\mathbf{z}_{2}\right)  \right\Vert
^{2}+c^{2}t^{2}\right)  \right]  }{\left[  \left\Vert \left(  \mathbf{z}%
_{1}-\mathbf{z}_{2}\right)  \right\Vert ^{2}+c^{2}t^{2}\right]  }%
\varphi\left(  \mathbf{z}_{2}\right)  d\mathbf{z}_{2} \tag{52}%
\end{equation}
where the order of integration was interchanged. We can easily see from
expression (52) that the solution is non-local(MacDonald's function of second
order) and does not coincide with the solution to the same problem where the
square root has been eliminated by reparametrization and introducing
constraints. It is important to note here the following:

i) from the semi-group representation of the radical operator we see that is
not the same to operate with the square root Hamiltonian as that with its
square or other power of this operator: \textit{the states under which the
Hamiltonian operates are sensible to the power of such Hamiltonian} .

ii) from the practical point of view the explicit determination of the
functions (states) $\varphi\left(  \mathbf{z}\right)  $ can carry several
troubles in any specific physical problems. In a future paper [28] we can give
a detailed study of this problem in different physical contexts.

\section{Relation with the relativistic Schr\"{o}dinger equation:
compatibility conditions and probability currents}

\bigskip Looking at formula (25) the new Hamiltonian operates as ($g_{\mu\nu
}=(+---)$)%
\[
\sqrt{m^{2}-\mathcal{P}_{0}\mathcal{P}^{0}-\left(  \mathcal{P}_{i}%
\mathcal{P}^{i}+\frac{1}{a}\Pi^{\alpha}\Pi_{\alpha}-\frac{1}{a^{\ast}}%
\Pi^{\overset{.}{\alpha}}\Pi_{\overset{.}{\alpha}}\right)  }\left\vert
\Psi\right\rangle =0
\]
for instance, the action of the radical operator is%
\begin{equation}
\left\{  \left[  m^{2}-\mathcal{P}_{0}\mathcal{P}^{0}-\left(  \mathcal{P}%
_{i}\mathcal{P}^{i}+\frac{1}{a}\Pi^{\alpha}\Pi_{\alpha}-\frac{1}{a^{\ast}}%
\Pi^{\overset{.}{\alpha}}\Pi_{\overset{.}{\alpha}}\right)  \right]  _{\beta
}^{\gamma}\left(  \Psi L_{\gamma}\right)  \Psi\right\}  ^{1/2}=0 \tag{53}%
\end{equation}
that seems as a parabosonic supersymmetric version of the relativistic
Schr\"{o}dinger-De Broglie equation. In the next paragraph we will see that
this equation corresponds to the family of equations given firstly by E.
Majorana [33] and P. A. M. Dirac [15], and in its para-bosonic version by
Sudarshan, N. Mukunda and C. C. Chiang in 1981[24].

We can see from the above expression that if we put the (super)momenta
together in the $\square$ operator, we obtain a more suitable equation in
order to compute the currents as in the Fock-Klein-Gordon case%
\begin{equation}
\left[  \left(  \square+m^{2}\right)  _{\beta}^{\alpha}\left(  \Psi L_{\alpha
}\Psi\right)  \right]  ^{1/2}=\left[  \left(  \square+m^{2}\right)  _{\beta
}^{\alpha}\Phi_{\alpha}\right]  ^{1/2}=0 \tag{54}%
\end{equation}
now eliminating the exponent $1/2$ and taking the Hermitian conjugation to
equation we have
\begin{align}
\left[  \left(  \square+m^{2}\right)  _{\beta}^{\alpha}\Psi^{\dagger}\left(
L_{\alpha}^{\dagger}\Psi^{\dagger}\right)  \right]  ^{1/2}  &  =0\tag{55}\\
\left[  \left(  \square+m^{2}\right)  _{\beta}^{\alpha}\left(  \Psi^{\dagger
}L_{\alpha}^{\dagger}\Psi^{\dagger}\right)  \right]  ^{1/2}  &  =\left[
\left(  \square+m^{2}\right)  _{\beta}^{\alpha}\Phi_{\alpha}^{\dagger}\right]
^{1/2}=0\nonumber
\end{align}
Following the same procedure as Dirac in Ref.[15] we multiply the square of
expression (54) from the left side by $\Phi_{\alpha}^{\dagger}$ and multiply
the square of expression (55) from the left side by $\Phi_{\alpha}$,
integrating and subtracting the final expressions we obtain%
\begin{equation}
\Phi_{\alpha}^{\dagger}\square\Phi_{\beta}-\Phi_{\alpha}\square\Phi_{\beta
}^{\dag}=0 \tag{56}%
\end{equation}
Using the relations: $\Phi^{\dagger}\square\Phi=\partial_{\mu}\left(
\Phi^{\dagger}\partial^{\mu}\Phi\right)  -\partial_{\mu}\Phi^{\dagger}%
\partial^{\mu}\Phi;\ \Phi\square\Phi^{\dagger}=\partial_{\mu}\left(
\Phi\partial^{\mu}\Phi^{\dagger}\right)  -\partial_{\mu}\Phi\partial^{\mu}%
\Phi^{\dagger}$ in expression (56) the current for the square states
$\Phi_{\alpha}$ is%
\begin{equation}
\partial_{\mu}\left(  \Phi^{\alpha}\partial^{\mu}\Phi_{\alpha}^{\dagger}%
-\Phi^{\alpha\dagger}\partial^{\mu}\Phi_{\alpha}\right)  =0=-\partial_{\mu
}j^{\mu} \tag{57}%
\end{equation}
with $j^{\mu}\left(  x\right)  \equiv-i\left[  \Phi^{\alpha}\partial^{\mu}%
\Phi_{\alpha}^{\dagger}-\Phi^{\alpha\dagger}\partial^{\mu}\Phi_{\alpha
}\right]  $.

If we suppose that a link between the relativistic Schr\"{o}dinger equation
(53) and our new Hamiltonian $\mathcal{H}$ holds, the relation with the
quartionic states is the following%
\[
i\overset{.}{\Psi}=E\Psi
\]
Squaring the above expression and having account as $\mathcal{H}$ operates
over $\Psi$ and $\Psi^{\dagger}$, we can easily obtain
\begin{align*}
-\overset{.}{\Phi_{\beta}}  &  =E^{2}\Phi_{\beta}\\
\overset{.}{\Phi_{\beta}^{\dagger}}  &  =E^{2}\Phi_{\beta}^{\dagger}%
\end{align*}
that into the explicit expression for $j_{0}\left(  x\right)  $ permits us to
analyze the positivity of this component of the current for the square states
$\Phi_{\alpha}$%
\begin{equation}
j_{0}\left(  x\right)  =2E^{2}\Phi^{\alpha\dagger}\Phi_{\alpha} \tag{58}%
\end{equation}
As we have been obtain from expression (58) $j_{0}\left(  x\right)  $ for the
square states $\Phi_{\alpha}$ is positively definite because the energy \ $E$
appears squared.

Now in order to find the current vector for the para-Bose states $\Psi$ we
proceed analogically as above for the states $\Phi_{\alpha}$ arriving to%
\[
\Psi^{\dagger}\square\Psi-\Psi\square\Psi^{\dagger}=0
\]
as we expected because these square root states obey the square root operator
equation and, for instance, also obey the equation for the squared operator
(the inverse is not true in general). In fact, in some references in the
literature the authors don't take care of the fact we can pass to the equation
with the square root Klein-Gordon operator to its squared traditional version
operating both on the same state but not the inverse (see e.g. ref.[34]). The
correct form to do this is as follows: if we start with%
\begin{equation}
\sqrt{\left(  -\Delta+m^{2}\right)  }\Psi=i\partial_{t}\Psi\tag{59}%
\end{equation}
the relation with any pseudo-differential operator $A$ is%
\[
A\Psi=\sqrt{\left(  -\Delta+m^{2}\right)  }\Psi=i\partial_{t}\Psi
\]%
\begin{align*}
AA\Psi &  =iA\partial_{t}\Psi=A\sqrt{\left(  -\Delta+m^{2}\right)  }\Psi\\
&  =\left(  -\Delta+m^{2}\right)  \Psi
\end{align*}
This happens clearly because $\Psi$ obeys (59). Finally the current for the
quartionic states that we were looking for is
\[
\partial_{\mu}\left[  \left(  \Psi\partial^{\mu}\Psi^{\dagger}\right)
-\left(  \Psi^{\dagger}\partial^{\mu}\Psi\right)  \right]  =0=-\partial_{\mu
}j^{\mu}%
\]%
\begin{equation}
j^{\mu}\left(  x\right)  \equiv-i\left[  \Psi\partial^{\mu}\Psi^{\dagger}%
-\Psi^{\dagger}\partial^{\mu}\Psi\right]  \tag{60}%
\end{equation}
Is not difficult to see that in this case from expression (60) the zero
component of the current is not positively definite given explicitly by%
\[
j_{0}\left(  x\right)  =2E\Psi^{\dagger}\Psi
\]
The compatibility condition, as usual, is given by the follow expression%
\begin{equation}
\left[  \tau_{\alpha},\tau_{\beta}\right]  \Psi=0 \tag{61}%
\end{equation}
where we defined $\tau_{\beta}\equiv\left[  \left(  \square+m^{2}\right)
_{\beta}^{\alpha}\left(  \Psi L_{\alpha}\right)  \right]  ^{1/2}$. After a
little algebra and using expression (61) we arrive to%
\begin{equation}
\left[  \left(  \square+m^{2}\right)  _{\alpha}^{\delta}\left(  \square
+m^{2}\right)  _{\beta}^{\gamma}\epsilon_{\delta\gamma}\right]  ^{1/2}\Psi=0
\tag{62}%
\end{equation}
It is good to remember here that eq.(53) describes a \textit{free} particle in
a N=1 superspace and the term of interaction appears from the supersymmetry
between the bosonic and fermionic fields. The last expression shows that our
equation (53) is absolutely compatible and consistent because its character
fermionic coming from the supersymmetric part, and for instance not necessary
to introduce any extra term in order to include spin. It is well known, that
it terms (put "by hand" in equations containing a second order derivatives)
destroy the compatibility condition leading to the impossibility of including
interactions [35].

\section{Relativistic wave equation}

Following the arguments given in the precedent paragraphs, it is natural to
propose the following form for a square root of the second order
supersymmetric wave equation%
\begin{equation}
\left\{  \left[  m^{2}-\mathcal{P}_{0}\mathcal{P}^{0}-\left(  \mathcal{P}%
_{i}\mathcal{P}^{i}+\frac{1}{a}\Pi^{\alpha}\Pi_{\alpha}-\frac{1}{a^{\ast}}%
\Pi^{\overset{.}{\alpha}}\Pi_{\overset{.}{\alpha}}\right)  \right]  _{\beta
}^{\alpha}\left(  \Psi L_{\alpha}\right)  \Psi\right\}  ^{1/2}=0 \tag{63}%
\end{equation}
In order to reduce the expression (63) to the simplest form it is necessary
pass to the center of mass of the system and redefining the variables as:
\[%
\begin{array}
[c]{ccc}%
t\rightarrow\left(  aa^{\ast}\right)  ^{-1/2}t,\ \ \ \  & \theta\rightarrow
a^{-1/2}\theta,\ \ \ \  & \overline{\theta}\rightarrow a^{-1/2}\overline
{\theta}%
\end{array}
\]
we obtain the following expression%
\[
\left\{  \left[  \left\vert a\right\vert ^{2}\partial_{0}^{2}+\frac{1}%
{4}\left(  \partial_{\eta}-\partial_{\xi}+i\ \partial_{0}\left(  \sigma
^{0}\right)  \xi\right)  ^{2}-\right.  \right.
\]%
\begin{equation}
\left.  \left.  -\frac{1}{4}\left(  \partial_{\eta}+\partial_{\xi}%
+i\ \partial_{0}\left(  \sigma^{0}\right)  \xi\right)  ^{2}+m^{2}\right]
_{\beta}^{\alpha}\Phi_{\alpha}\right\}  ^{1/2}=0 \tag{64}%
\end{equation}
where%
\begin{equation}%
\begin{array}
[c]{ccc}%
\eta\equiv\left(  \overline{\theta}+\theta\right)  , & \ \ \xi\equiv\left(
\overline{\theta}-\theta\right)  , &
\ \ \ \ \ \ \ \ \ and\ \ \ \ \ \ \ \partial_{0}\left(  \sigma^{0}\right)
_{\alpha\overset{.}{\beta}}\left(  \overline{\theta}^{\overset{.}{\beta}%
}-\theta^{\alpha}\right)  \equiv\partial_{0}\left(  \sigma^{0}\right)  \xi
\end{array}
\tag{65}%
\end{equation}
The trick that we used above [21,36,37] is based on the observation that the
expression (63) has similar form that the equation for an electron in constant
electromagnetic field (with $\mathcal{P}_{\mu}\left(  \sigma^{\mu}\right)
_{\alpha\overset{.}{\beta}}$ as the constant electric field). Imposing the
condition $\partial_{\eta}\Phi_{\alpha}=0\ \Rightarrow\Phi_{\alpha}\left(
\xi\right)  $; the "square" of the solution eigenfunction of eq.(63) takes the
form
\begin{equation}
\Phi_{\gamma}\left(  t\right)  =e^{A\left(  t\right)  +\xi\varrho\left(
t\right)  }\Phi_{\gamma}\left(  0\right)  \tag{66}%
\end{equation}
with $\varrho\left(  t\right)  =\phi_{\alpha}+\overline{\chi}_{\overset
{.}{\alpha}}$(i.e.chiral plus anti-chiral parts). The system of equations for
$A\left(  t\right)  $ and $\varrho\left(  t\right)  $ that we are looking for,
is easily obtained inserting the expression (66) in the eq.(64)%
\[
\left\vert a\right\vert ^{2}\ddot{A}+m^{2}=0
\]%
\[
\overset{..}{\overline{\chi}}_{\overset{.}{\alpha}}-i\frac{\omega}{2}\left(
\sigma^{0}\right)  _{\ \overset{.}{\alpha}}^{\alpha}\ \phi_{\alpha}=0
\]%
\[
-\overset{..}{\phi}_{\alpha}+i\frac{\omega}{2}\left(  \sigma^{0}\right)
_{\alpha}^{\overset{.}{\ \beta}}\ \overline{\chi}_{\overset{.}{\beta}}=0
\]
The above system can be solved given us the following result%
\begin{equation}
A=-\left(  \frac{m}{\left\vert a\right\vert }\right)  ^{2}t^{2}+c_{1}%
t+c_{2}\ ;\ \ c_{1},c_{2}\in\mathbb{C} \tag{67}%
\end{equation}
and%
\begin{equation}
\phi_{\alpha}=\overset{\circ}{\phi}_{\alpha}\left(  \alpha e^{i\omega
t/2}+\beta e^{-i\omega t/2}\right)  +\frac{2i}{\omega}\left(  \sigma
^{0}\right)  _{\alpha}^{\overset{.}{\ \beta}}\ \overline{Z}_{\overset{.}%
{\beta}} \tag{68}%
\end{equation}%
\begin{equation}
\overline{\chi}_{\overset{.}{\alpha}}=\left(  \sigma^{0}\right)
_{\ \overset{.}{\alpha}}^{\alpha}\ \overset{\circ}{\phi}_{\alpha}\left(
\alpha e^{i\omega t/2}-\beta e^{-i\omega t/2}\right)  +\frac{2i}{\omega
}\left(  \sigma^{0}\right)  _{\ \overset{.}{\alpha}}^{\alpha}\ Z_{\alpha}
\tag{69}%
\end{equation}
where $\overset{\circ}{\phi}_{\alpha},\ Z_{\alpha}$ and $\overline
{Z}_{\overset{.}{\beta}}$ are constant spinors. The superfield solution for
the square states that we are looking for, have the following form%
\begin{equation}
\Phi_{\gamma}\left(  t\right)  =e^{-\left(  \frac{m}{\left\vert a\right\vert
}\right)  ^{2}t^{2}+c_{1}t+c_{2}}e^{\xi\varrho\left(  t\right)  }\Phi_{\gamma
}\left(  0\right)  \tag{70}%
\end{equation}
with
\begin{equation}
\varrho\left(  t\right)  =\overset{\circ}{\phi}_{\alpha}\left[  \left(  \alpha
e^{i\omega t/2}+\beta e^{-i\omega t/2}\right)  -\left(  \sigma^{0}\right)
_{\ \overset{.}{\alpha}}^{\alpha}\ \left(  \alpha e^{i\omega t/2}-\beta
e^{-i\omega t/2}\right)  \right]  +\frac{2i}{\omega}\left[  \left(  \sigma
^{0}\right)  _{\alpha}^{\overset{.}{\ \beta}}\ \overline{Z}_{\overset{.}%
{\beta}}+\left(  \sigma^{0}\right)  _{\ \overset{.}{\alpha}}^{\alpha
}\ Z_{\alpha}\right]  \tag{71}%
\end{equation}
and%
\begin{equation}
\Phi_{\gamma}\left(  0\right)  =\left\langle \Psi\left(  0\right)  \right\vert
L_{\gamma}\left\vert \Psi\left(  0\right)  \right\rangle \tag{72}%
\end{equation}
that is nothing more that the mean value of $L_{\gamma}$ between the states
$\left\vert \Psi\right\rangle $ in the initial time, where the subalgebra is
the Heisenberg-Weyl algebra (with generators $a,$ $a^{+}$ and$\left(
n+\frac{1}{2}\right)  $). As we have pointed out in Section V, the states
$\left\vert \Psi\right\rangle $\ span all the Hilbert space and for instance,
we can not obtain useful information from the point of view of the topology of
the group manifold then, about the spin.

The dynamics of the square root fields, in the representation that we are
interested in, can be simplified considering these fields as coherent states
in the sense that are eigenstates of $a^{2}$%

\begin{align}
\left\vert \Psi_{1/4}\left(  0,\xi,q\right)  \right\rangle  &  =\overset
{+\infty}{\underset{k=0}{\sum}}f_{2k}\left(  0,\xi\right)  \left\vert
2k\right\rangle =\overset{+\infty}{\underset{k=0}{\sum}}f_{2k}\left(
0,\xi\right)  \frac{\left(  a^{\dag}\right)  ^{2k}}{\sqrt{\left(  2k\right)
!}}\left\vert 0\right\rangle \tag{73}\\
\left\vert \Psi_{3/4}\left(  0,\xi,q\right)  \right\rangle  &  =\overset
{+\infty}{\underset{k=0}{\sum}}f_{2k+1}\left(  0,\xi\right)  \left\vert
2k+1\right\rangle =\overset{+\infty}{\underset{k=0}{\sum}}f_{2k+1}\left(
0,\xi\right)  \frac{\left(  a^{\dagger}\right)  ^{2k+1}}{\sqrt{\left(
2k+1\right)  !}}\left\vert 0\right\rangle \nonumber
\end{align}
From a technical point of view these states are a one-mode squeezed states
constructed by the action of the generators of the SU(1,1) group over the
vacuum. For simplicity, we will take all normalization and fermionic
dependence or possible CS fermionic realization, into the functions $f\left(
\xi\right)  $. Explicitly at t=0
\begin{equation}%
\begin{array}
[c]{c}%
\left\vert \Psi_{1/4}\left(  0,\xi,q\right)  \right\rangle =f\left(
\xi\right)  \left\vert \alpha_{+}\right\rangle \\
\left\vert \Psi_{3/4}\left(  0,\xi,q\right)  \right\rangle =f\left(
\xi\right)  \left\vert \alpha_{-}\right\rangle
\end{array}
\tag{74}%
\end{equation}
where $\left\vert \alpha_{\pm}\right\rangle $ are the CS basic states in the
subspaces $\lambda=\frac{1}{4}$ and $\lambda$ =$\frac{3}{4}$ of the full
Hilbert space. From expression (70) and expressions (42) we obtain%
\begin{equation}
\Phi_{\alpha}\left(  t,\lambda\right)  =\left\langle \Psi_{\lambda}\left(
t\right)  \right\vert \mathbb{L}_{\alpha}\left\vert \Psi_{\lambda}\left(
t\right)  \right\rangle =e^{-\left(  \frac{m}{\left\vert a\right\vert
}\right)  ^{2}t^{2}+c_{1}t+c_{2}}e^{\xi\varrho\left(  t\right)  }\left\langle
\Psi_{\lambda}\left(  0\right)  \right\vert \left(
\begin{array}
[c]{c}%
a^{2}\\
\left(  a^{+}\right)  ^{2}%
\end{array}
\right)  _{\alpha}\left\vert \Psi_{\lambda}\left(  0\right)  \right\rangle
\tag{75}%
\end{equation}%
\begin{equation}
\Phi_{\alpha}\left(  t,\lambda\right)  =e^{-\left(  \frac{m}{\left\vert
a\right\vert }\right)  ^{2}t^{2}+c_{1}t+c_{2}}e^{\xi\varrho\left(  t\right)
}\left\vert f\left(  \xi\right)  \right\vert ^{2}\left(
\begin{array}
[c]{c}%
\alpha_{\lambda}^{2}\\
\alpha_{\lambda}^{\ast2}%
\end{array}
\right)  _{\alpha} \tag{76}%
\end{equation}
where $\lambda$ label the helicity or the spanned subspace (e.g. $\pm$). The
"square root" states solution of the expression (63) take the following form%
\begin{equation}
\Psi_{\lambda}=e^{-\frac{1}{2}\left[  \left(  \frac{m}{\left\vert a\right\vert
}\right)  ^{2}t^{2}+c_{1}t+c_{2}\right]  }e^{\frac{\xi\varrho\left(  t\right)
}{2}}\left\vert f\left(  \xi\right)  \right\vert \left(
\begin{array}
[c]{c}%
\alpha\\
\alpha^{\ast}%
\end{array}
\right)  _{\lambda} \tag{77}%
\end{equation}
where $\lambda=1/4,3/4.$ Notice the difference with the case in which we used
the HW realization for the states $\Psi$%
\begin{equation}
\left\vert \Psi\right\rangle =\frac{f\left(  \xi\right)  }{2}\left(
\left\vert \alpha_{+}\right\rangle +\left\vert \alpha_{-}\right\rangle
\right)  =f\left(  \xi\right)  \left\vert \alpha\right\rangle \tag{78}%
\end{equation}
where, however, the linear combination of the states $\left\vert \alpha
_{+}\right\rangle $ and $\left\vert \alpha_{-}\right\rangle $ span now the
full Hilbert space being the corresponding $\lambda$ to this CS basis
$\lambda=\frac{1}{2}$ .The "square" state at t=0 are%
\begin{align}
\Phi_{\alpha}\left(  0\right)   &  =\left\langle \Psi\left(  0\right)
\right\vert L_{\alpha}\left\vert \Psi\left(  0\right)  \right\rangle
=\left\langle \Psi\left(  0\right)  \right\vert \left(
\begin{array}
[c]{c}%
a\\
a^{+}%
\end{array}
\right)  _{\alpha}\left\vert \Psi\left(  0\right)  \right\rangle \tag{79}\\
&  =f^{\ast}\left(  \xi\right)  f\left(  \xi\right)  \left(
\begin{array}
[c]{c}%
\alpha\\
\alpha^{\ast}%
\end{array}
\right)  _{\alpha}\nonumber
\end{align}
The square state at time t%
\begin{equation}
\Phi_{\gamma}\left(  t\right)  =e^{-\left(  \frac{m}{\left\vert a\right\vert
}\right)  ^{2}t^{2}+c_{1}^{\prime}t+c_{2}^{\prime}}e^{\xi\varrho\left(
t\right)  }\left\vert f\left(  \xi\right)  \right\vert ^{2}\left(
\begin{array}
[c]{c}%
\alpha\\
\alpha^{\ast}%
\end{array}
\right)  _{\alpha} \tag{80}%
\end{equation}
And the "square root" solution becomes now to%
\begin{equation}
\Psi\left(  t\right)  =e^{-\frac{1}{2}\left[  \left(  \frac{m}{\left\vert
a\right\vert }\right)  ^{2}t^{2}+c_{1}^{\prime}t+c_{2}^{\prime}\right]
}e^{\frac{\xi\varrho\left(  t\right)  }{2}}\left\vert f\left(  \xi\right)
\right\vert \left(
\begin{array}
[c]{c}%
\alpha^{1/2}\\
\alpha^{\ast1/2}%
\end{array}
\right)  \tag{81}%
\end{equation}
We can see the change in the solutions from the choice in the representation
of the Hilbert space. The algebra (topological information of the group
manifold) is "mapped" over the spinors solutions through the eigenvalues
$\alpha$ and $\alpha^{\ast}$. Notice that the constants $c_{1}^{\prime}$
$c_{2}^{\prime}$ in the exponential functions \ in expressions (80) and (81)
differ from the $c_{1}$ and $c_{2}$ in (76) and (77), because these
exponential functions of the Gaussian type come from the action of a unitary
operator over the respective CS basic states in each representation ($h_{3}$
or HW). These constants can be easily determined as functions of the frequency
$\omega$ as in ref. [5] for the Schrodinger equation. A detailed analysis of
this point and the other type of solutions will be given elsewhere [28].

About the possible algebras that contain an SU(1,1) as subgroup that can lead
or explain the fermionic factors of type $e^{\frac{\xi\varrho\left(  t\right)
}{2}}\left\vert f\left(  \xi\right)  \right\vert $ in the solutions are 2
subgroups that are strong candidates [38]: the supergroup $OSP\left(
2,2\right)  $[2] and the supergroup $OSP\left(  1/2,\mathbb{R}\right)  $[39].
In the case of the $OSP\left(  2,2\right)  $ we have bosonic and fermionic
realizations and the CS and SS can be constructed from the general procedure
given by M. Nieto et al. in Refs.[1-3]. On the other hand, the $OSP\left(
1/2,\mathbb{R}\right)  $ realization is more "economic", the number of
generators is minor than in the $OSP\left(  2,2\right)  $ case and the
realization is bosonic: the $K_{\pm}$ and $K_{0}$ generators operate over the
bose states and the HW algebra given by $a$ and $a^{+}$ operate over the
fermionic part. In this case the CS and the SS that can be constructed are
eigenstates of the displacement and squeezed operators respectively but they
cannot minimize the dispersion of the quadratic Casimir operator, so that they
are not minimum uncertainty states.

The important point to remark here is that when we describe from \ the mostly
geometrical grounds any physical system through SU(1,1) CS or SS, the orbits
will appear as the intersections of curves that represent constant-energy
surfaces, with one sheet of a two sheeted hyperboloid- the curved phase space
of SU(1,1) or Lobachevsky plane- in the space of averaged algebra generators.
In the specific case treated in this paper, the group containing the SU(1,1)
as subgroup linear and bilinear functions of the algebra generators can
factorize operators as the Hamiltonian or the Casimir operator (when averaged
with respect to group CS or SS), defining corresponding curves in the averaged
algebra space. If we notice that the validity of the Ehrenfest's theorem for
CS (SS) implies that, if the exact dynamics is confined to the SU(1,1)
hyperboloid, it necessarily coincides with the variational motion, the
variational motion that comes from the Euler-Lagrange equations for the
lagrangian%
\[
\mathcal{L}=\left\langle z\right\vert i\frac{\widehat{\partial}}{\partial
t}-\widehat{H}\left\vert z\right\rangle
\]
will be different if $\left\vert z\right\rangle =\left\vert \alpha
\right\rangle $ or $z=\left\vert \alpha_{\pm}\right\rangle $, as is evident to
see. It is interesting to note also that similar picture holds in the context
of the pseudospin \ SU(1,1) dynamics in the frame of the mean field
approximation induced by the variational principle on nonlinear Hamiltonians [40].

\section{Concluding remarks}

In this work the problem of the physical interpretation of the square root
quantum operators\ and possible relation with the TDHO\ and coherent and
squeezed states was analyzed considering the simple model of superparticle of
Volkov and Pashnev [9]. Besides the extension and clarification of the results
of our previous works [19], of which we have already made mention in the
Introduction, we can summarize as follows:

i) the Fock construction for these fractional or \textquotedblright square
root states\textquotedblright\ was proposed, explicitly constructed and
compared with the Fock construction given in the reference[9] for the
superparticle model with the Hamiltonian in standard form;

ii) we have shown that, in contrast to [9], the only states that the square
root Hamiltonian can operate with correspond to the representations with the
lowest weights $\lambda=\frac{1}{4}$ and $\lambda=\frac{3}{4}$ ;

iii) there are four possible (non-trivial) fractional representations for the
group decomposition of the spin structure from the square root Hamiltonian,
instead of (1/2,0) and (0,1/2) as the case when the Hamiltonian is quadratic
in momentum (e.g. Ref. [9]) as a consequence of the geometrical Hamiltonian
taken in its natural square root form and the Sannikov-Dirac oscillator
representation for the generators of the Lorentz group SO(3,1);

now we make this research complete with the following new results:

iv) The relation between the structure of the Hilbert space of the states, the
spin content of the sub-Hilbert spaces and the CS and SS realization of the
physical states was established for the particular model presented here.

iv) We construct explicitly from the theory of semi-group the analytical
representation of the radical operator in the N=1 superspace and we see that
it is not the same to operate with the square root Hamiltonian as that with
its square or other power of this operator from the point of view of the
spectrum of the physical states: \textit{the states under which the
Hamiltonian operates are sensible to the power of such Hamiltonian} .

v) If expression constructed in iv) gives a closed representation for the
radical operator, from the practical point of view the explicit determination
of the functions (states) $\varphi\left(  \mathbf{z}\right)  $ can carry
several troubles in any specific physical problems.

vi) The relation between the relativistic Schr\"{o}dinger equation and other
type of equations that involve variables with fractional spin\ and the model
analyzed here was established and discussed.

vii) As for the Klein-Gordon equation, the conserved currents for the
"square-root" states (para-fields) and for the square states were explicitly
computed and analyzed. The component zero of the current is linearly dependent
on the energy E in the para-field case and for the "square" state the
dependence on the energy is quadratic .

viii) The compatibility conditions were analyzed and the consistency of the
proposed equation was established. The explanation of this consistency and the
relation with the free dynamics and the supersymmetry of the model was given.

ix) New wave equation is proposed and explicitly solved for the time-dependent
case. As for the TDHO the physical states are realized on the CS and SS basis,
and the link between the topology of the (super)-group manifold and the
obtained solution from the algebraic and group theoretical point of view was
discussed and analyzed.

It is interesting to see that the results presented here for the superparticle
are in complete agreement with the results, symmetry group and discussions for
non-supersymmetric examples given in references [41,42,43], where group and
geometrical quantization was used. This fact gives a high degree of
reliability of our method of quantization and the correct interpretation of
the radical Hamiltonian operator. It is clear that the ordinary Canonical
method of quantization fails when the reparametrization procedure affects the
power of the starting Hamiltonian modifying inexorably the obtained spectrum
of the physical states [see e.g. [42,43]]. For instance, we conclude that
quantically it is not the same to operate with the square root Hamiltonian as
that with its square or other power of this operator because the obtained
states (mass spectrum, spin) under which the Hamiltonian operates are sensible
to the power of such Hamiltonian.; and seeing that the lowest weights of the
states under the square root Hamiltonian can operate, and because not concrete
action is known to describe particles with fractional statistics,
superparticle relativistic actions as of [9] can be good geometrical and
natural candidates to describe quartionic states [13,16,17,18] (semions).

\section{Acknowledgements}

I am very thankful to Professors John Klauder for his advisements, E. C. G.
Sudarshan for his interest demonstrated in this work in a private
communication and in particular to N. Mukunda for his interest to put in this
research and give me several remarks and references on the problem of
consistency of the positive energy wave equations with para-fields variables.
I appreciate deeply Professor Tepper Gill's efforts to clarify to me many
concepts on the correct description of the quantum systems, Hamiltonian
formulation and the relation with the proper energy of these systems. I am
very grateful to the Directorate of JINR, in particular of the Bogoliubov
Laboratory of Theoretical Physics, for their hospitality and support.
Note added: when this paper was alredy finished, the Professors M. Plyushcay and S. Gavrilov put our attention on references [44] and [45]
where the canonical quantization was performed in non-commutative formulation and in the ordinary space respectively 
We are very acknowledge to M. Plyushcay and S. Gavrilov for your very valuable references.

\section{References}

1. B. W. Fatyga \textit{et al.}, \textit{Phys. Rev. }D \textbf{43}, 1403
(1991), and references therein.

2. V. A. Kostelecky \textit{et al.}, \textit{Phys. Rev. }A \textbf{48}, 1045
(1993), and references therein.

3. V. A. Kostelecky \textit{et al.}, \textit{Phys. Rev.} A \textbf{48}, 951
(1993), and references therein.

4. V. Ermakov, \textit{Univ. Izv. Kiev} Serie III \textbf{9}, 1 (1880).

5. K. Husimi, \textit{Prog. Theor. Phys.}\textbf{ 9}, 381, (1953).

6. J. R. Klauder and B. S. Skagerstam:\textit{Coherent States} (World
Scientific, Singapore, 1985).

7. M. D. Shelby \textit{et al.}, \textit{Phys. Rev. Lett.} \textbf{57}, 691 (1986).

8. C. M. Caves et al., \textit{Rev. Mod. Phys.}\textbf{ 52}, 341 (1980).

9. A.I. Pashnev and D. V. Volkov: Supersymmetric lagrangian for particles in
proper time. \textit{Teor. Mat. Fiz.}, Tom. 44, No.\textbf{ 3}, 321 (1980) [in Russian].

10. R. Casalbuoni: The Classical Mechanics for Bose-Fermi
Systems.\ \textit{Nuovo. Cim.}, Vol. \textbf{33A}, N. 3, 389 (1976).

11. R. Casalbuoni: Relatively and supersymmetries. \textit{Phys. Lett}.
\textbf{62}B, 49 (1976).

12. C. Lanczos: \textit{Variational Principles in Mechanics,} (Mir, 1965), pp.
408 (Russian version).

13. Yu. P. Stepanovsky: On massless fields and relativistic wave equations.
\textit{Nucl. Phys.} B (Proc. Suppl.), \textbf{102-103}, 407 (2001).

14. S. S. Sannikov: Non-compact symmetry group of a quantum oscillator.
\textit{Zh.E.T.F.} \textbf{49}, 1913 (1965), [in Russian].

15. P. A. M. Dirac: A positive-energy relativistic wave equation.
\textit{Proc. Roy. Soc. }\textbf{A 322}, 435 (1971).

16. D. P. Sorokin and D. V. Volkov: (Anti) commuting spinors and
supersymmetric dynamics of semions. \textit{Nucl. Phys.} \textbf{B409}, 547 (1993).

17. D. P. Sorokin: The Heisenberg algebra and spin. \textit{Fortschr. Phys.}
\textbf{50}, 724 (2002).

18. D. V. Volkov: Quartions in relativistic field theories. \textit{Piz'ma
Zh.E.T.F.} \textbf{49}, 473 (1989), [in Russian].

19. D. J. Cirilo-Lombardo, \textit{Rom. Journal of Physics} \textbf{7-8}, Vol
50, 875, (2005); \textit{Elem. Particles and Nucl. Lett.}\textbf{6, }Vol. 3,
416 (2006); \textit{Hadronic J. }\textbf{29,} 355 (2006).

20. A. P. Akulov and D. V. Volkov: Is the neutrino a Goldstone particle?
\textit{Phys. Lett.} \textbf{46B}, 109 (1973).

21. A. S. Bakai and Yu. P. Stepanovsky: \textit{Adiabatic Invariants
}(\textquotedblright Naukova Dumka\textquotedblright, Kiev, 1981), pp. 65 [in Russian].

22. J. Sucher: Relativistic invariance and the Square-Root Klein-Gordon
equation. \textit{J. Math. Phys.}\textbf{\ 4}, 17 (1963), and references therein.

23. S. Schweber: \textit{An Introduction to Relativistic Quantum Field Theory}
(Row, Peterson and Co., Evanston, Illinois, 1964), pp. 56.

24. E. C. G. Sudarshan \textit{et al.}: Dirac positive energy wave equation
with para-Bose internal variables, \textit{Phys. Rev.} \textbf{D 25}, 3237 (1982).

25. E. C. G. Sudarshan and N. Mukunda, \textit{Phys. Rev.} D \textbf{1}, 571 (1970).

26. N. Mukunda \textit{et al.}, \textit{J. Math. Phys. }\textbf{21}, 2386 (1980).

27. T.L. Gill and W.W. Zachary, \textit{J.Phys.A: Math. and General}
\textbf{38, }2479, (2005).

28. D. J. Cirilo-Lombardo: work in preparation.

29. A. Rogers: A global theory of supermanifolds, \textit{J. Math. Phys.}
\textbf{21}, 1352 (1980).

30. B. De Witt: \textit{Supermanifolds} (Cambridge University Press,
Cambridge, 1984).

31. R. F. Picken and K. Sundermeyer : Integration on Supermanifolds and a
Generalized Cartan Calculus, \textit{Comm. Math. Phys.} \textbf{102}, 585 (1986).

32. K. Yosida, \textit{Functional Analysis} (Springer, New York, 2nd. edition, 1968).

33. E.Majorana, \textit{Nuovo Cim.} \textbf{9}, 335 (1932).

34. C. L\"{a}mmerzahl, \textit{J. Math. Phys.} \textbf{34}, 3918 (1993).

35. R. P. Feynman and M. Gellman, \textit{Phys. Rev.} \textbf{109}, 193 (1958).

36. D. J. Cirilo-Lombardo and Yu. P. Stepanovsky, \textit{Problems on Atomic
Science and Technology }\textbf{6}, 182 (2001).

37. Akhiezer A. I. and Beretsetsky, V. B.: \textit{Quantum Electrodynamics},
p.p. 432 (Nauka, Moscow, 1981).

38. I. Bars and M. Gunaydin, \textit{Commun. Math. Phys.} \textbf{91}, 31 (1983).

39. Le-Man Kuang and Xing Chen, \textit{J.Phys.A: Math. and General}
\textbf{27, }L119, (1993).

40. D. M. Jezek and H. S. Hernandez, \textit{Phys. Rev.} A \textbf{42}, 96
(1990), and references therein.

41. M. Lachieze-Rey: On three quantization methods for a particle on
hyperboloid, gr-qc/0503060, (2005).

42. R. Delbourgo: A square root of the harmonic oscillator, hep-th/9503056, (1995).

43. E. Elizalde: On the concept of determinant for the differential operators
of quantum physics. \textit{JHEP} \textbf{07}, 015 (1999).

44. Peter A. Horvathy (Tours U., CNRS) , Mikhail S. Plyushchay, Mauricio Valenzuela (Santiago de Chile U.) . Oct 2006. 15pp. 
 Nucl.Phys.B768:247-262,2007. 

45. D. Gitman and S. Gavrilov,Int. J. of Mod. Phys. A 15
(2000) 4499-4538.

\bigskip

\bigskip
\end{document}